\documentclass[twocolumn]{aastex62}
\usepackage{amssymb,multirow,morefloats,amsmath}
\usepackage{threeparttable}
\usepackage{longtable}
\usepackage{epstopdf}
\usepackage{graphicx}
\accepted{September 18, 2019}
\submitjournal{ApJ}

\shorttitle{XRB metallicity dependence at z$\sim$2}
\shortauthors{Fornasini et al.}

\newcommand\T{\rule{0pt}{2.6ex}}       
\newcommand\B{\rule[-1.2ex]{0pt}{0pt}} 
\defcitealias{mineo12}{M12}
\defcitealias{lehmer10}{L10}
\defcitealias{lehmer16}{L16}
\defcitealias{fragos13b}{F13}
\defcitealias{madau17}{M17}
\defcitealias{aird17}{A17}
\defcitealias{brorby16}{B16}

\begin{document}

\title{The MOSDEF Survey: The Metallicity Dependence of X-ray Binary Populations at $z\sim2$}

\correspondingauthor{Francesca M. Fornasini}
\email{francesca.fornasini@cfa.harvard.edu}

\author[0000-0002-9286-9963]{Francesca M. Fornasini}
\affil{Center for Astrophysics $|$ Harvard \& Smithsonian, 60 Garden Street, Cambridge, MA 02138, USA}

\author[0000-0002-7613-9872]{Mariska Kriek}
\affil{Astronomy Department, University of California, Berkeley, CA 94720, USA}

\author[0000-0003-4792-9119]{Ryan L. Sanders}
\affil{Department of Physics \& Astronomy, University of California, Los Angeles, 430 Portola Plaza, Los Angeles, CA 90095, USA}

\author[0000-0003-4702-7561]{Irene Shivaei}
\affil{Department of Astronomy/Steward Observatory, 933 North Cherry Avenue, Room N204, Tucson, AZ, 85721-0065, USA}
\affil{Hubble Fellow}

\author{Francesca Civano}
\affil{Center for Astrophysics $|$ Harvard \& Smithsonian, 60 Garden Street, Cambridge, MA 02138, USA}

\author[0000-0001-9687-4973]{Naveen A. Reddy}
\affil{Department of Physics \& Astronomy, University of California, Riverside, 900 University Avenue, Riverside, CA 92521, USA}

\author{Alice E. Shapley}
\affil{Department of Physics \& Astronomy, University of California, Los Angeles, 430 Portola Plaza, Los Angeles, CA 90095, USA}

\author[0000-0002-2583-5894]{Alison L. Coil}
\affil{Center for Astrophysics and Space Sciences, University of California, San Diego, 9500 Gilman Drive, La Jolla, CA 92093-0424, USA}

\author{Bahram Mobasher}
\affil{Department of Physics \& Astronomy, University of California, Riverside, 900 University Avenue, Riverside, CA 92521, USA}

\author[0000-0002-4935-9511]{Brian Siana}
\affil{Department of Physics \& Astronomy, University of California, Riverside, 900 University Avenue, Riverside, CA 92521, USA}

\author[0000-0003-1908-8463]{James Aird}
\affil{Department of Physics \& Astronomy, University of Leicester, University Road, Leicester LE1 7RJ, UK}

\author[0000-0001-6004-9728]{Mojegan Azadi}
\affil{Center for Astrophysics $|$ Harvard \& Smithsonian, 60 Garden Street, Cambridge, MA 02138, USA}

\author[0000-0003-3559-5270]{William R. Freeman}
\affil{Department of Physics \& Astronomy, University of California, Riverside, 900 University Avenue, Riverside, CA 92521, USA}

\author[0000-0002-9393-6507]{Gene C. K. Leung}
\affil{Center for Astrophysics and Space Sciences, University of California, San Diego, 9500 Gilman Drive, La Jolla, CA 92093-0424, USA}

\author[0000-0002-0108-4176]{Sedona H. Price}
\affil{Max-Planck-Institut f\"{u}r extraterrestrische Physik, Postfach 1312, Garching, D-85741, Germany}

\author{Tara Fetherolf}
\affil{Department of Physics \& Astronomy, University of California, Riverside, 900 University Avenue, Riverside, CA 92521, USA}

\author{Tom Zick}
\affil{Astronomy Department, University of California, Berkeley, CA 94720, USA}

\author[0000-0001-6813-875X]{Guillermo Barro}
\affil{Department of Physics, University of the Pacific, 3601 Pacific Avenue, Stockton, CA 95211, USA}

\begin{abstract}
Population synthesis models predict that high-mass X-ray binary (HMXB) populations produced in low metallicity environments should be more X-ray luminous, a trend supported by studies of nearby galaxies.  This trend may be responsible for the observed increase of the X-ray luminosity ($L_{\mathrm{X}}$) per star formation rate (SFR) with redshift due to the decrease of metallicity ($Z$) at fixed stellar mass as a function of redshift.  To test this hypothesis, we use a sample of 79 $z\sim2$ star-forming galaxies with oxygen abundance measurements from the MOSDEF survey, which obtained rest-frame optical spectra for $\sim1500$ galaxies in the CANDELS fields at $1.37<z<3.80$.  Using \textit{Chandra} data from the AEGIS-X Deep, Deep Field North, and Deep Field South surveys, we stack the X-ray data at the galaxy locations in bins of redshift and $Z$ because the galaxies are too faint to be individually detected.  In agreement with previous studies, the average $L_{\mathrm{X}}$/SFR of our $z\sim2$ galaxy sample is enhanced by $\approx0.4-0.8$ dex relative to local HMXB $L_{\mathrm{X}}$-SFR scaling relations.  Splitting our sample by $Z$, we find that $L_{\mathrm{X}}$/SFR and $Z$ are anti-correlated with 97\% confidence.  This observed $Z$ dependence for HMXB-dominated galaxies is consistent both with the local $L_{\mathrm{X}}$-SFR-$Z$ relation and a subset of population synthesis models.  Although the statistical significance of the observed trends is weak due to the low X-ray statistics, these results constitute the first direct evidence connecting the redshift evolution of $L_{\mathrm{X}}$/SFR and the $Z$ dependence of HMXBs.  
\end{abstract}


\section{Introduction} 
\label{sec:intro}
Studies of nearby star-forming galaxies have established that the integrated X-ray luminosity ($L_{\mathrm{X}}$) of high-mass X-ray binaries (HMXBs) in a galaxy is linearly correlated with its star formation rate (SFR; \citealt{ranalli03}; \citealt{grimm03}; \citealt{persic04}; \citealt{gilfanov04a}; \citealt{lehmer10}; \citealt{mineo12}). This $L_{\mathrm{X}}$-SFR correlation exists because of the young ages and short lifetimes of HMXBs, which consist of a black hole (BH) or neutron star (NS) accreting material from a high-mass ($M>8M_{\odot}$) stellar companion.  It is estimated that HMXBs form just $\sim$4-40 Myr after a starburst and remain X-ray active only for $\sim10$ Myr (\citealt{iben95}; \citealt{bodaghee12c}; \citealt{antoniou16}).  \par
Several studies have found that the normalization of the $L_{\mathrm{X}}$-SFR relation evolves with redshift ($z$), increasing by about 0.5 dex in $L_{\mathrm{X}}$ at fixed SFR between $z=0-2$ (\citealt{basu13a}; \citealt{lehmer16}; \citealt{aird17}).  On the contrary, \citet{cowie12} found no redshift evolution of $L_{\mathrm{X}}$/SFR.  However, \citet{basu13a} argued that this apparent lack of evolution results from the fact that \citet{cowie12} did not correct their SFR proxy, UV luminosity, for dust extinction, and \citet{kaaret14} suggested that the anomalous \citet{cowie12} results may be attributed to their adoption of a spectral model that was not steep enough at hard X-ray energies.  It has been suggested that the redshift evolution of $L_{\mathrm{X}}$/SFR is driven by the metallicity ($Z$) dependence of HMXB populations and the fact that, on average, HMXBs at higher redshift have lower stellar metallicities.   \par
Over the past decade, binary population synthesis studies have investigated the effects of metallicity on HMXB evolution.  The winds of main-sequence high-mass stars are primarily driven by radiation pressure on atomic lines.  Since high-mass stars primarily emit in the UV, and metals have far more UV atomic lines than H or He, the strength of their stellar winds is determined by their stellar metallicity.  
As a result, higher $Z$ stars experience higher mass loss rates, losing more mass during the course of their lifetimes than lower $Z$ stars.  Therefore, the compact objects in low $Z$ binaries are expected to be more massive than ones produced by stars of similar initial mass in higher $Z$ binaries (\citealt{belczynski04}; \citealt{dray06}; \citealt{fragos13a}).  Another effect of the weaker winds of lower $Z$ stars is that less angular momentum is lost from the binary, resulting in a larger fraction of HMXBs in which accretion occurs via Roche lobe overflow \citep{linden10}.  Thus, lower $Z$ HMXB populations are expected to contain larger fractions of Roche lobe overflow BH HMXBs, which are typically more luminous than wind-fed NS HMXBs.  There is a general consensus that larger populations of luminous HMXBs exist in lower $Z$ environments, although the strength of this trend varies between studies.  Studies predict that $L_{\mathrm{X}}$/SFR may increase by factor of 2 to 10 between $Z_{\odot}$ and 0.1 $Z_{\odot}$ (\citealt{linden10}; \citealt{fragos13b}).  \par
In addition to informing models of binary stellar evolution, constraining the $Z$ dependence of HMXB populations can yield insight into possible formation channels for the heavy BH binaries discovered by gravitational wave observatories \citep{abbott16}.  Such BH binaries are thought to have evolved either from HMXBs in a low $Z$ environment \citep{belczynski16} or through dynamical formation in dense stellar clusters \citep{rodriguez16}.  Constraining the $Z$ dependence of HMXBs is also critical for understanding their contribution to the X-ray heating of the intergalactic medium during the epoch of reionization when the Universe was extremely metal-poor (\citealt{mirabel11}; \citealt{madau17}), and informing models of the 21 cm power spectrum (e.g. \citealt{parsons14}).  Furthermore, these constraints are important to accurately estimate the HMXB contamination to X-ray based searches for intermediate mass black holes (\citealt{mezcua17}).  Finally, properly calibrating for the $Z$ dependence of HMXBs can improve the reliability of $L_{\mathrm{X}}$ as a SFR indicator \citep{brorby16}.   
\par
There is increasing observational evidence that a larger number of HMXBs, especially ultra-luminous X-ray sources (ULXs, $L_X\gtrsim10^{39}$ erg s$^{-1}$), per unit SFR exist in nearby low $Z$ galaxies (\citealt{mapelli11}; \citealt{kaaret11}; \citealt{prestwich13}; \citealt{basu13b}; \citealt{brorby14}; \citealt{douna15}).  The enhanced number of bright HMXBs cannot be accounted for by stochasticity and suggests that at very low metallicities (12+log(O/H) $< 8.0$), the production rate of HMXBs is approximately 10 times higher than in solar metallicity (12+log(O/H)=8.69) galaxies (\citealt{brorby14}; \citealt{douna15}).  Using a compilation of measurements for 49 galaxies from the literature spanning $7.0<$12+log(O/H)$<9.0$, \citet{brorby16} (hereafter \citetalias{brorby16}) parametrize the $L_X$-SFR-$Z$ correlation as:
\begin{equation}
\mathrm{log}\left(\frac{L_X/SFR}{\mathrm{erg\hspace{4pt} s^{-1}/(M_{\odot}\mathrm{\hspace{4pt}yr^{-1})}}}\right) = b\times(12+\mathrm{log(O/H)}-8.69)+c 
\end{equation}
where the best-fitting parameters are $b=-0.59\pm0.13$ and $c=39.49\pm0.09$.  While this relation may be biased due to the mixture of sample selections for different galaxy samples taken from the literature, it provides the first observational benchmark of the $L_{\mathrm{X}}$-SFR-$Z$ relation at $z=0$.  However, it has not yet been shown that the $Z$ dependence of HMXBs is the underlying cause of the observed redshift evolution of $L_{\mathrm{X}}$/SFR.  \par
We present the results of an X-ray stacking study of $z\sim2$ star-forming galaxies drawn from the MOSFIRE Deep Evolution Field (MOSDEF) survey whose goal is to test this hypothesis observationally.  The MOSDEF survey obtained rest-frame optical spectra for roughly 1500 galaxies at $1.4<z<3.8$ in CANDELS fields, which have been observed to deep limits with the \textit{Chandra X-ray Observatory}; the combination of a large sample of high-redshift galaxies with robust $Z$ measurements and deep X-ray data is what makes the study of the connection between the redshift evolution and $Z$ dependence of HMXBs possible for the first time.  In \S \ref{sec:mosdef}, we describe the MOSDEF survey and the measurement of galaxy properties.  \S\ref{sec:chandra} describes the \textit{Chandra} X-ray data and catalogs used in this study.  Sample selection and our X-ray stacking analysis are detailed in \S\ref{sec:sample} and \S\ref{sec:stacking}, respectively.  In \S\ref{sec:results}, we discuss our measurement of the $Z$ dependence of $L_X$/SFR at $z\sim2$ and compare it to the local $L_X$-SFR-$Z$ relation and theoretical models.  Our conclusions are presented in \S\ref{sec:conclusions}.  Throughout this work, we assume a cosmology with $\Omega_m = 0.3$, $\Omega_{\Lambda} = 0.7$, and $h = 0.7$ and adopt the solar abundances from \citet{asplund09} ($Z_{\odot}=0.0142$, 12+log(O/H)$_{\odot}=8.69$).

\section{The MOSDEF Survey}
\label{sec:mosdef}
Our $z\sim2$ galaxy sample is selected from the MOSDEF survey \citep{kriek15}.  This survey obtained moderate-resolution (R = 3000--3650) rest-frame optical spectra for $\sim1500$ $H$-band selected galaxies using the MOSFIRE multi-object near-IR spectrograph \citep{mclean12} on the 10-meter Keck I telescope.  
MOSDEF targets are located in the CANDELS fields, where extensive multi-wavelength coverage is available (\citealt{grogin11}; \citealt{koekemoer11}).  Possible MOSDEF target objects were selected from the 3D-HST photometric and spectroscopic catalogs (\citealt{skelton14}; \citealt{momcheva16}) to magnitude limits of $H= 24.0$, $H = 24.5$, and $H = 25.0$ for the low (1.37 $\leq z \leq$ 1.70), middle (2.09 $\leq z \leq$ 2.61), and high (2.95 $\leq z \leq$ 3.80) redshift intervals, respectively.  These magnitude limits roughly correspond to a lower mass limit of $\sim10^9M_{\odot}$.  \par
The three redshift intervals were chosen to maximize coverage of strong rest-frame optical emission lines such that they fall within atmospheric transmission windows.  Hereafter we will refer to these redshift intervals as $z\sim1.5$, $z\sim2.3$, and $z\sim3.4$, respectively, and collectively refer to the galaxies in the two lowest redshift intervals as the $z\sim2$ sample.

\subsection{MOSDEF Data Reduction}
The MOSFIRE spectra were reduced using a custom automated pipeline which performs flat-fielding, subtracts sky background, cleans cosmic rays, rectifies the frames, combines all individual exposures for a given source, and calibrates the flux (see \citealt{kriek15} for details).  
Slit-loss corrections were determined by modeling the HST $H$-band light distribution of galaxies and calculating the amount of light passing through the slit, as detailed in \citet{kriek15} and \citet{reddy15}.
One-dimensional science and error spectra were optimally extracted based on the algorithm of \citet{horne86} (see the Appendix in \citealt{freeman19} for details). \par

\subsection{Emission lines fluxes and spectroscopic redshifts}
Emission-line fluxes were measured by fitting Gaussian line profiles on top of a linear continuum to the one-dimensional spectra (\citealt{kriek15}; \citealt{reddy15}).  The H$\alpha$ and H$\beta$ emission line fluxes were corrected for Balmer absorption using best-fit SED models, as described in \citet{reddy18}.  Flux uncertainties were estimated by performing 1,000 Monte Carlo realizations of the spectrum of each object perturbed by its error spectrum and refitting the line profiles; the average line fluxes and dispersions were measured from the resulting line flux distributions.  Spectroscopic redshifts were measured using the centroids of the highest signal-to-noise ratio (S/N) emission lines, typically H$\alpha$ or [O{\small III}] $\lambda$5007.  In total, the MOSDEF survey obtained spectroscopic redshifts for roughly 1300 objects, including galaxies that were specifically targeted and those that were serendipitously observed.

\subsection{SED-derived M$_*$ and SFR}
\label{sec:sed}
Stellar masses were estimated by modeling the available photometric data \citep{skelton14} for each galaxy with the spectral energy distribution (SED) fitting program FAST \citep{kriek09}, adopting the MOSDEF-measured spectroscopic redshift for each galaxy.  The photometric data span rest-frame UV to near-IR wavelengths for $z\sim2$ galaxies.  We used the stellar population synthesis models of \citet{conroy09}, assumed a \citet{chabrier03} IMF, adopted the \citet{calzetti00} dust attenuation curve, and parametrized star-formation histories using delayed exponentially declining models of the form SFR$(t) \propto te^{-t/\tau}$, where $t$ is the time since the onset of star formation and $\tau$ is the characteristic star formation timescale. For each galaxy, the best-fitting model was found through $\chi^2$ minimization, and confidence intervals for all free parameters were calculated from the distributions of 500 Monte Carlo simulations which perturbed the input photometric data points and repeated the SED fitting procedure.

Since our goal is to study the relationship between $L_{\mathrm{X}}$/SFR and $Z$, we explored the effect that SED-fitting assumptions can have on the derived SFR, particularly those assumptions that are $Z$-dependent (see \S\ref{sec:sfrindicator}).  Therefore, we also used the results of SED fits from \citet{reddy18}, which use the \citet{bruzual03} (hereafter BC03) stellar population models and vary the assumed dust attenuation curve (\citealt{calzetti00} or SMC from \citealt{gordon03}) and the stellar metallicity ($Z=0.02$ or $Z=0.004$).  These SED fits assume constant SF histories, which have been shown to be appropriate for typical ($L^*$) galaxies at $z\gtrsim1.5$ by previous studies \citep{reddy12}.  Prior to SED-fitting, the photometry was corrected for the contribution from the strongest emission lines in the MOSFIRE spectra, including [O II], H$\beta$, [O III], and H$\alpha$.  

\subsection{H$\alpha$ SFR}
\label{sec:hasfr}

SFRs were also derived from dust-corrected H$\alpha$ luminosities.  H$\alpha$ SFRs are sensitive to SF on shorter timescales and subject to partly different systematics than SED-derived SFRs.  H$\alpha$ luminosities were corrected for dust attenuation using the absorption-corrected Balmer decrement (H$\alpha$/H$\beta$) as described in \citet{reddy15} and \citet{shivaei15}.  These corrections assume the \citet{cardelli89} extinction curve, which Reddy et al. (in prep) find is consistent with the nebular reddening curve of MOSDEF $z\sim2$ galaxies. The dust-corrected H$\alpha$ luminosities were converted into SFRs using the calibration of \citet{hao11} assuming a \citet{chabrier03} IMF (conversion factor of $4.634\times10^{-42} M_{\odot}$ yr$^{-1}$ erg$^{-1}$ s).  H$\alpha$-derived SFRs are only calculated for galaxies in which both H$\alpha$ and H$\beta$ are detected with S/N $\geq3$.  \par
Since this conversion factor depends on $Z$, \citet{reddy18} derive an alternative conversion factor that is more appropriate for the MOSDEF sample based on the BC03 $Z=0.004$ (0.28 $Z_{\odot}$) model.  This conversion factor is $3.236\times10^{-42} M_{\odot}$ yr$^{-1}$ erg$^{-1}$ s.     \par
For our default SFR measurements, we adopt H$\alpha$ SFRs with a $Z$-dependent correction, wherein for galaxies with 12+log(O/H)$>$8.3, we apply the H$\alpha$ luminosity conversion factor appropriate for $Z=0.02$ from \citet{hao11}, and for galaxies with lower O/H, we adopt the conversion factor for $Z=0.004$ from \citet{reddy18}.  Although we think these SFR measurements are the most robust, nonetheless in \S\ref{sec:sfrindicator}, we discuss the impact that our choice of SFR indicator has on our results.  \par
While the SED-derived SFRs may not fully account for dust-obscured star formation because they are based on fitting rest-frame UV to near-IR data, the H$\alpha$ SFRs are found to be in good agreement with UV+IR SFRs.  \citet{shivaei16} compared the H$\alpha$ SFRs of 17 MOSDEF galaxies detected by \textit{Spitzer} MIPS and \textit{Herschel} with SED-derived SFRs based on the UV to far-IR bands, and found strong agreement with 0.17 dex of scatter but no systematic biases.

\subsection{Metallicity}

The gas-phase metallicity of a galaxy is derived from the fluxes of emission lines originating from gas in H{\small II} regions found near sites of recent star formation.  Thus, the gas-phase oxygen abundance (O/H) is often used as a proxy for the stellar metallicity of the young stellar population of a galaxy, including its HMXBs.  In order to facilitate the comparison of our results to the local $L_{\mathrm{X}}$-SFR-$Z$ relation, we adopt the same O/H indicator as \citet{brorby16}, namely O3N2 (log(([O{\small III}]$\lambda$5007/H$\beta$)/([N{\small II}]$\lambda$6584/H$\alpha$))).  We use the calibration of \citet{pettini04}, which is based on a sample of H{\small II} regions most of which have direct electron temperature measurements.  This calibration is:
\begin{equation}
12+\mathrm{log(O/H)} = 8.73 - 0.32 \times \mathrm{O3N2}
\end{equation}
We require that the four emission lines used for the O3N2 indicator are not significantly affected by nearby skylines or too close to the edge of the spectrum to measure the line flux reliably.  If one or more of the emission lines required to calculate the O3N2 flux ratio was not detected with S/N $\geq3$, then a 3$\sigma$ upper limit on the line flux was computed and used to calculate an upper or lower limit on O/H.  
\par
Even though we are using the same O/H indicator as \citet{brorby16}, it is possible that the O3N2 indicator evolves with redshift (\citealt{shapley15}; \citealt{sanders16a}) and that chemical abundances in galaxies at $z\sim2$ differ from the solar pattern (\citealt{steidel16}; \citealt{sanders19}), affecting the relationship between gas-phase O/H and stellar metallicity.  We discuss the systematic effects on our results due to these effects in \S\ref{sec:systematics}.  

\subsection{AGN Identification}
\label{sec:agn}
In order to study the X-ray binary (XRB) emission from MOSDEF galaxies, it is important to remove all known active galactic nuclei (AGN) from our sample.  We identify AGN using diagnostics in multiple wavelength bands as detailed in \citet{coil15}, \citet{azadi17}, and \citet{leung17}.  The AGN identification criteria are summarized below and the possible impact of AGN contamination is discussed in \S\ref{sec:systematics}.   

\subsubsection{X-ray AGN}
All MOSDEF galaxies with \textit{Chandra} counterparts were classified as X-ray AGN.  \citet{coil15} matched \textit{Chandra} sources detected by \texttt{wavdetect} with a false probability threshold $<4\times10^{-6}$ in at least one of four energy bands (0.5--7, 0.5--2, 2--7, and 4--7~keV) to likely multi-wavelength counterparts using the likelihood ratio method described in \citet{nandra15}; then the closest matches within 1$^{\prime\prime}$ to these counterparts were found in the 3D-HST catalogs used for MOSDEF target selection.  The X-ray detected MOSDEF galaxies at $z>1.3$ have high rest-frame 2-10~keV luminosities of $L_{\mathrm{X}}>10^{41.5}$ erg s$^{-1}$ indicative of AGN emission.\footnote{In fact, all but one of the X-ray detected MOSDEF galaxies at $z>1.3$ have $L_{\mathrm{X}}>10^{42.5}$ erg s$^{-1}$.}  
\par

\subsubsection{IR AGN}
Since X-ray photons are absorbed at very high column densities ($N_{\mathrm{H}}\gtrsim10^{24}$ cm$^{-2}$), X-ray surveys can miss the most heavily obscured AGN.  In these obscured sources, the high-energy AGN emission is processed by dust and re-radiated at mid-infrared (MIR) wavelengths.  This phenomenon makes it is possible to identify these obscured AGN based on their MIR colors.  We select IR AGN using data from the Infrared Array Camera (IRAC; \citealt{fazio04}) on \textit{Spitzer} reported in the 3D-HST catalogs \citep{skelton14} and the IRAC color selection criteria defined by \citet{donley12}.  \par

\subsubsection{Optical AGN}
Optical diagnostics such as the ``BPT diagram'' (\citealt{baldwin81}; \citealt{veilleux87}) can be used to identify AGN via their enhanced ratios of nebular emission lines [O{\small III}]$\lambda5008$/H$\beta$ and [N{\small II}]$\lambda6584$/H$\alpha$.  However, since these diagnostics are based on the narrow components of emission lines, more detailed fitting of the H$\alpha$, H$\beta$, [O{\small III}], and [N{\small II}] emission lines is required to properly decompose the broad and narrow line components.  As described in more detail in \citet{azadi17} and \citet{leung17}, the emission lines were fit with up to three Gaussian components: a narrow, a broad, and a blueshifted component representing outflows.  
The broad and outflow components were only accepted if they resulted in an improved fit at $>99$\% confidence.  Galaxies with significant broad lines were identified as optical AGN. \par
Using only the narrow line components, we placed galaxies on the BPT diagram.  For this study, we flagged as an optical AGN any galaxy with log([N{\small II}]/H$\alpha$)$>-0.3$ and any galaxy falling above the \citet{kauffmann03} line in the BPT diagram.  Not all galaxies above the \citet{kauffmann03} line are expected to be AGN, especially at $z\gtrsim2$ where galaxies are found to be offset to higher [N{\small II}]/H$\alpha$ values at fixed [O{\small III}]/H$\beta$ (\citealt{masters14}; \citealt{shapley15}; \citealt{sanders16a}; \citealt{strom17}).  However, we choose to be conservative in our sample selection since even low-luminosity AGN emission may contaminate our measurements of X-ray luminosity, SFR, and O/H.

\section{\textit{Chandra} Extragalactic Surveys}
\label{sec:chandra}
The \textit{Chandra X-ray Observatory} has performed several deep extragalactic surveys.  For this study, we use the \textit{Chandra} ACIS imaging in the \textit{Chandra} AEGIS-X Deep, Deep Field North (CDF-N), and Deep Field South (CDF-S) fields.  These fields have the deepest X-ray exposures, permitting the most complete identification and removal of X-ray AGN.  The exposure depths reached in these fields is 7~Ms in CDF-S, 2~Ms in CDF-N, and 800~ks in AEGIS-XD (\citealt{alexander03}; \citealt{nandra15}; \citealt{luo17}).
 The corresponding flux limits (over $>50$\% of the survey area) in the 0.5--2~keV band reached by these surveys are $5\times10^{-17}$, $1.2\times10^{-16}$, and $2\times10^{-16}$~erg~cm$^{-2}$~s$^{-1}$, respectively, which correspond to 2--10~keV rest-frame X-ray luminosities of $1.2\times10^{42}$, $3.5\times10^{42}$, $5.8\times10^{42}$~erg~s$^{-1}$ at $z\sim2$ assuming a power-law spectrum with a photon index of $\Gamma=2.0$.  \par

\subsection{\textit{Chandra} data processing}
The \textit{Chandra} data from the AEGIS-XD and CDF-N fields were processed as described in \citet{laird09}, \citet{rangel13}, \citet{nandra15}, and \citet{aird15}, and we made use of the publicly available \textit{Chandra} mosaic images and exposure maps of the CDF-S field produced as described in \citet{luo17}.  The data were processed using the CIAO analysis software v4.1.2 and v4.8 for the former and latter data sets respectively.  The data processing procedures applied to all three datasets are very similar and are briefly summarized below with full details provided in \citet{nandra15} and \citet{luo17}.  \par
Each observation was cleaned and calibrated using standard CIAO algorithms. For each observation, the \textit{Chandra} wavelet source detection algorithm \texttt{wavdetect} was run on the 0.5--7~keV band image with a detection threshold of $10^{-6}$.  The astrometry of individual observations was improved by using the CIAO tool \texttt{reproject\_aspect} to minimize the offsets between \texttt{wavdetect} sources and counterparts in the Canada-France-Hawaii Telescope Legacy survey (CFHTLS) $i$-band catalog \citep{gwyn12} for AEGIS-XD, the $r$-band Hawaii HDFN catalog \citet{capak04} for CDF-N, and the Taiwan ECDFS Near-Infrared Survey (TENIS) K$s$-band catalog \citep{hsieh12} for CDF-S. 

\subsection{Chandra data products and catalogs}
For each individual observation, event files, images, exposure maps, and PSF maps of the 90\% encircled energy fraction (EEF) as calculated by the MARX simulator were created for the 0.5--7, 0.5--2, and 2--7~keV bands.  The exposure maps provide the exposure multiplied by the effective collecting area at each pixel location; they are weighted for a power-law spectrum with $\Gamma=1.4$, the photon index of the cosmic X-ray background (\citealt{gruber99}; \citealt{ajello08}; \citealt{cappelluti17}).\footnote{Adopting a different value of $\Gamma$ in the range from 1.0 to 2.0 would change the exposure map values by $<10\%$.}  The event files, images, exposure maps, and PSF maps for each field were merged together into the mosaics used in our X-ray stacking analysis (see \S\ref{sec:stacking}).  \par
We compared the astrometric frame of these final mosaics to the astrometry of the 3D-HST catalogs from which the MOSDEF galaxy positions are determined.  For each X-ray detected counterpart of a MOSDEF galaxy with $>40$ net counts in the $0.5-7$ keV band, we calculated the $(x,y)$ positional offset between its coordinates from the 3D-HST catalog and its centroid coordinates measured using the \texttt{gcntrd} IDL program.  In the AEGIS, GOODS-S, and GOODS-N fields, there are 9, 6, and 23 such counterparts to MOSDEF galaxies.  For each field, we then determined the average $x$ and $y$ offset, and found them to be $<1$ pixel in all cases.  We apply these positional shifts to ensure the best match between the \textit{Chandra} mosaics and the astrometric reference frame of MOSDEF galaxies.  However, we note that because the X-ray aperture regions used in our stacking analysis are at least 4 pixels in diameter, these small positional shifts do not significantly impact the derived X-ray properties of our stacks.  \par
In order to study the XRB emission of non-AGN MOSDEF galaxies, it is important to reduce not only contamination from MOSDEF AGN but also the contribution from nearby detected X-ray sources that are not associated with MOSDEF galaxies.  X-ray source catalogs for the CDF-S, CDF-N, and AEGIS-XD surveys are provided by \citet{luo17}, \citet{alexander03}, and \citet{nandra15}, respectively.  We use these catalogs to remove the contribution of detected \textit{Chandra} sources to our X-ray stacks as detailed in \S\ref{sec:stacking}. \par

\section{Galaxy Sample Selection}
\label{sec:sample}
Since our goal is to study HMXB emission as a function of $Z$ and SFR, we apply several selection criteria to select galaxies with reliably measured properties and to minimize contamination from other X-ray sources.  \par
Therefore, we excluded from our sample any MOSDEF galaxy that is identified as an AGN using the X-ray, IR, or optical criteria described in \S\ref{sec:agn}.   
The other sources that can contribute significantly to a galaxy's hard X-ray emission are low-mass X-ray binaries (LMXBs), whose X-ray emission is correlated with the stellar mass ($M_*$) of a galaxy (\citealt{gilfanov04}; \citealt{colbert04}).  Thus, the X-ray contribution of HMXBs relative to LMXBs is maximized in galaxies with high specific SFR (sSFR=SFR/$M_*$).  Studies of local galaxies find that galaxies with sSFR$>10^{-10}$~yr$^{-1}$ are HMXB-dominated \citep{lehmer10}, which is true of all galaxies in our MOSDEF sample.  However, the sSFR value at which galaxies transition from being LMXB-dominated to HMXB-dominated may increase with redshift (\citealt{lehmer16}).  Considering that this value for $z\sim2$ galaxies remains poorly constrained, we study how restricting our sample to different sSFR ranges affects our results.  Our MOSDEF galaxies span a sSFR range of $10^{-9.6}-10^{-8.1}$ yr$^{-1}$ with a median sSFR of $10^{-8.8}$ yr$^{-1}$.  \par
We required that galaxies have an O/H measurement or upper/lower limit based on the O3N2 indicator in order to facilitate comparison to local studies of the $Z$ dependence of HMXBs (\citealt{douna15}; \citealt{brorby16}).  In addition, we restricted our sample to galaxies with both H$\alpha$ and SED-derived SFRs.   
We further limited our galaxy sample to $M_*\geq10^{9.5} M_{\odot}$, because \citet{shivaei15} demonstrated that MOSDEF samples may be incomplete at lower $M_*$ due to a bias against young objects with small Balmer and 4000 $\mathrm{\AA}$ breaks.  Furthermore, the MOSDEF survey may not be complete in H$\alpha$ SFRs for $M_*<10^{9.5} M_{\odot}$, because we would expect that such low-mass galaxies scattering below the $M_*$-SFR relation \citep{sanders18} would fall below the $3\sigma$ H$\beta$ detection limit. 
\par  
As mentioned in \S\ref{sec:chandra}, even though the MOSDEF survey covers all the CANDELS fields, we only included galaxies from the GOODS-S \citep{giavalisco04}, GOODS-N \citep{giavalisco04}, and AEGIS \citep{davis07} fields; the \textit{Chandra} survey of the COSMOS field \citep{scoville07} is much shallower \citep{civano16}, and in the UKIDSS-UDS field \citep{lawrence07} only 34 galaxies were observed as a part of MOSDEF, just a few of which meet our selection criteria.  Furthermore, only galaxies in the $z\sim1.5$ and $z\sim2.3$ redshift intervals are used.  The galaxy sample at $z\sim3.4$ is too small to produce significant X-ray stacked detections and, at these high redshifts, the H$\alpha$ and [N{\small II}] emission lines move out of the near-infrared band, requiring different diagnostics to screen optical AGN and to measure $Z$ and SFR. \par
Finally, we imposed two additional restrictions in order to optimize the \textit{Chandra} stacking procedure described in \S\ref{sec:stacking}. These criteria are based on the size of the PSF at the galaxy position and proximity to other sources.\par

\begin{figure*}
\centering
\includegraphics[width=1.0\linewidth]{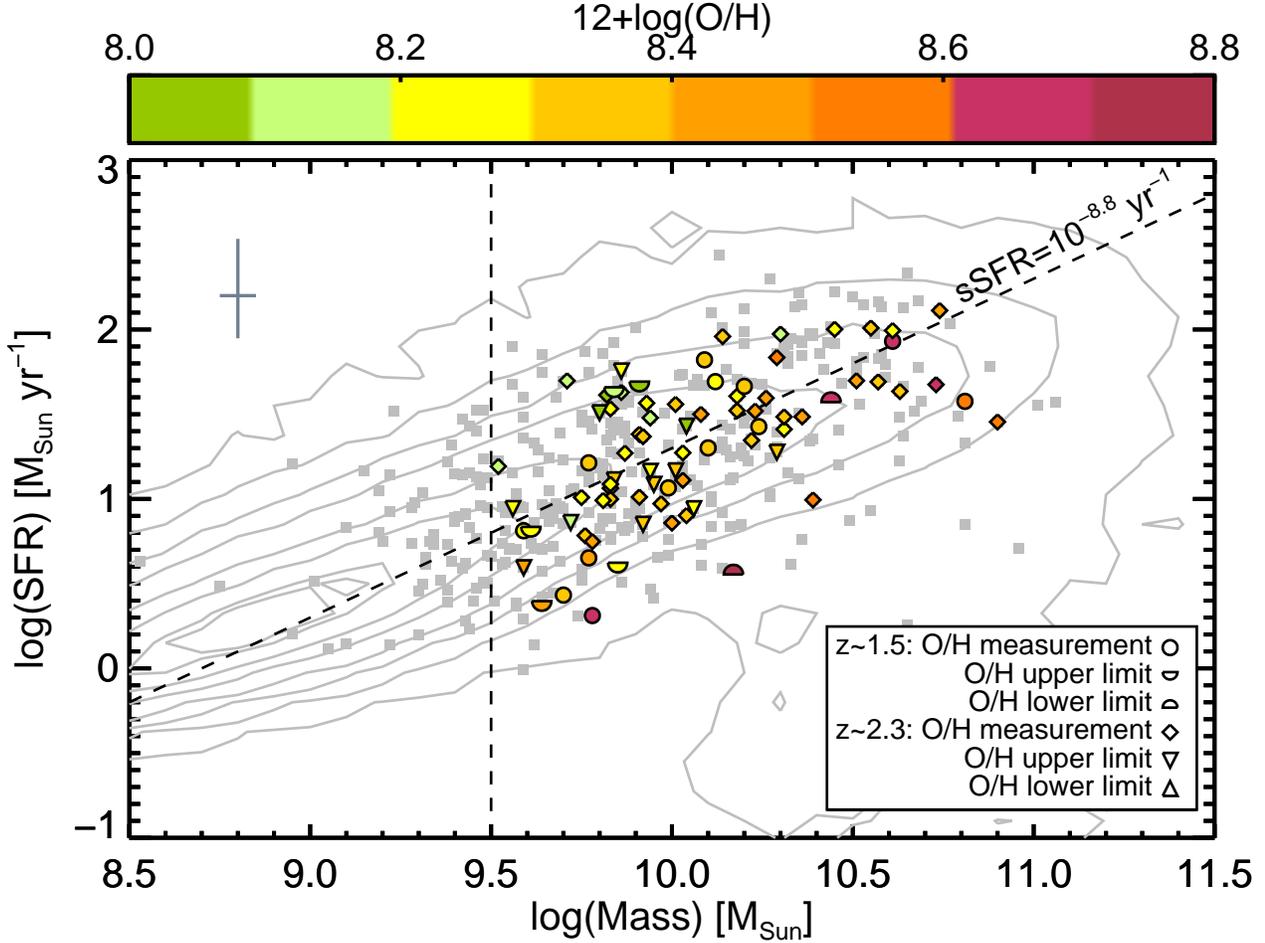}
\vspace{0.5in}
\caption{H$\alpha$ SFRs of MOSDEF galaxies versus $M_*$ derived from SED fitting.  The gray squares show all MOSDEF galaxies with H$\alpha$-derived SFRs and no AGN signatures.  The 79 galaxies used in our analysis are shown by symbols colored according to $Z$.  The circle, lower half circle, and upper half circle symbols represent galaxies at 1.37 $\leq z \leq$ 1.70 with O/H measurements, upper limits, or lower limits, respectively.   The diamond, downward triangle, and upward triangle symbols represent galaxies at 2.09 $\leq z \leq$ 2.61 with O/H measurements, upper limits, or lower limits, respectively.  The median 1$\sigma$ uncertainty in SFR and $M_*$ measurements is shown in the upper left corner.  The MOSDEF galaxies are well-distributed along the main sequence of star-forming galaxies shown by the gray contours, which are based on the distribution of galaxies with 1.37 $\leq z \leq$ 2.61 from the COSMOS field \citep{laigle16}. The vertical dashed line represents our $M_*$ selection threshold, while the diagonal dashed line represents our division of the sample into high and low sSFR galaxies (sSFR$=10^{-8.8}$ yr$^{-1}$).}
\label{fig:sfrmass}
\end{figure*}

\begin{figure}
\centering
\includegraphics[width=1.0\linewidth]{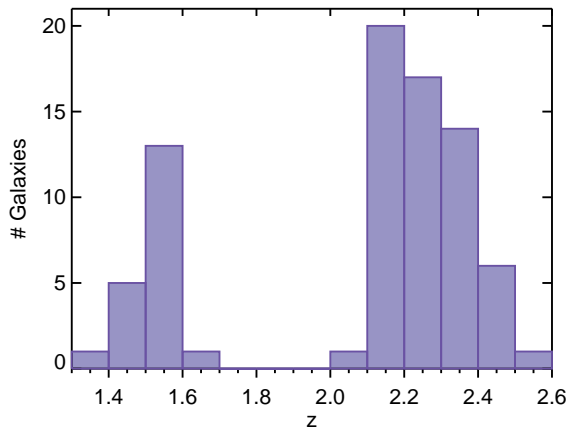}
\caption{The redshift distribution of the 79 galaxies used for our study.  The distribution is bimodal because MOSDEF targets were selected in specific redshift windows ($1.37\leq z\leq1.70$ and $2.09\leq z\leq2.61$) for which rest-frame optical strong emission lines fall within atmospheric transmission windows.  There are more galaxies with $z\sim2.3$ than $z\sim1.5$ as a result of the MOSDEF survey's targeting strategy \citep{kriek15}.}
\label{fig:redshift}
\end{figure}

\begin{figure}
\centering
\includegraphics[width=1.0\linewidth]{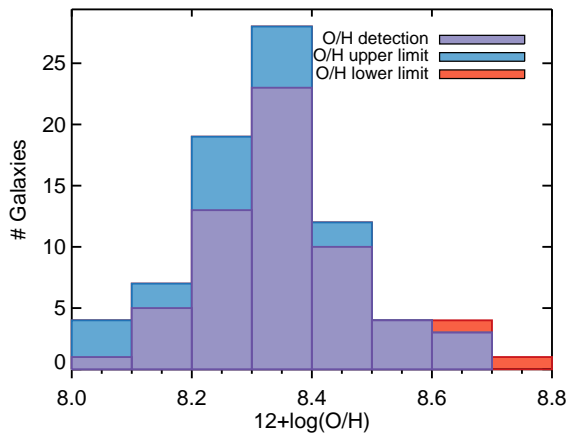}
\caption{The distribution of $Z$ of the galaxies in our sample, as traced by nebular O/H based on the O3N2 indicator.  O/H measurements (calculated when all relevant emission lines are significantly detected) are shown in purple, while O/H upper and lower limits are shown in blue and red, respectively.}
\label{fig:metaldist}
\end{figure}

Only 79 MOSDEF galaxies meet all our selection criteria.  Figure \ref{fig:sfrmass} displays the H$\alpha$ SFR versus stellar mass for the galaxies in our sample in points colored by the oxygen abundance and outlined in blue or black for $z\sim1.5$ and $z\sim2.3$, respectively.  The gray dots in this figure represent all MOSDEF galaxies with $1.4<z<2.7$ and H$\alpha$-derived SFRs, and the gray contours are based on the SED-derived SFRs and $M_*$ from a much larger sample of $\sim$160,000 galaxies in the COSMOS field with photometric redshifts in the same $z$ range \citep{laigle16}.  As can be seen, even though our sample is small, it is representative of star-forming galaxies at $z\sim2$.  \citet{sanders18} found evidence that a $M_*$-SFR-$Z$ relation exists in MOSDEF galaxies at $z\sim2.3$, a hint of which can be observed in Figure \ref{fig:sfrmass} as galaxies with higher SFR at fixed $M_*$ have lower $Z$.  As shown in Figure \ref{fig:redshift}, 20 galaxies are at $z\sim1.5$ and 59 are at $z\sim2.3$.  The $Z$ distribution of our sample is shown in Figure \ref{fig:metaldist}, with 12+log(O/H) measurements, 3$\sigma$ upper, and 3$\sigma$ lower limits indicated in different colors.  The $Z$ distribution is strongly peaked at 12+log(O/H) = 8.3-8.4.

\section{X-ray Stacking Analysis}
\label{sec:stacking}
The typical X-ray luminosities of normal (non-AGN) star-forming galaxies are $L_X<10^{42}$~erg~s$^{-1}$ in the rest-frame 2--10~keV band.\footnote{Low luminosity AGN and/or obscured AGN can also exhibit $L_X<10^{42}$~erg~s$^{-1}$.  We discuss possible contamination from such AGN in \S\ref{sec:systematics}.}  Since these luminosities fall below the sensitivity limits of the \textit{Chandra} extragalactic surveys at $z\sim2$, studying the X-ray emission of these galaxies requires stacking the X-ray data.  We developed an X-ray stacking technique that is similar to that used by other studies (e.g. \citealt{basu13a}; \citealt{rangel13}; \citealt{mezcua16}; \citealt{fornasini18}).  In order to achieve the highest sensitivity, we performed the stacking primarily using the 0.5--2~keV band because \textit{Chandra} has the highest effective area and best angular resolution at soft X-ray energies.  However, we also stacked the data in the 2-7~keV band in order to measure the hardness ratios for our stacks. \par
\subsection{X-ray photometry for individual sources}
For each of the galaxies in our sample, we defined source and background aperture regions.  Each source aperture was defined as a circular region centered on the galaxy position from the 3D-HST catalog with a radius equal to the 90\% ECF PSF radius ($r_{90}$).  The median $r_{90}$ of MOSDEF objects is $2.4^{\prime\prime}$, and their angular sizes are small enough that their galaxy-wide X-ray emission is consistent with a point source.  Each background aperture was defined as an annulus with an inner radius equal to 10$^{\prime\prime}$ and an outer radius equal to 30$^{\prime\prime}$.  \par
As mentioned in \ref{sec:sample}, we made some refinements to our galaxy sample based on X-ray criteria.  
In order to prevent contamination to our source or background count estimates from unassociated X-ray sources, we masked out from the images and exposure maps circular regions with a radius of 2$r_{90}$ at the positions of any X-ray detected sources in the CDF-S, CDF-N, and AEGIS-XD source catalogs (\citealt{alexander03}; \citealt{luo17}; \citealt{nandra15}), regardless of what energy band the source was detected in.  We removed from our sample any galaxy at a distance $<2r_{90}$ of an X-ray detected source.  We also excluded any galaxies at a distance $<r_{90}$ from another galaxy in the 3D-HST catalog with $H_{\mathrm{AB}}$ (F160W) magnitude $<24.0$, which is the magnitude limit approximately corresponding to $M_*=10^{9.5} M_{\odot}$ at $z\sim2.3$.  These sources were excluded to avoid contamination from neighboring galaxies with X-ray emission below the sensitivity threshold of the \textit{Chandra} surveys but potentially of comparable or greater luminosity as our MOSDEF galaxies.  We also removed four galaxies located in an area of diffuse X-ray emission in the CDF-N field.  Finally, to optimize the signal-to-noise ratio, we excluded galaxies located far off-axis in the \textit{Chandra} observations as the \textit{Chandra} PSF increases with off-axis angle; we find that the significance of our stacks is maximized by excluding sources with $r_{90}>3.5^{\prime\prime}$.   \par
We extracted the counts ($C_{\mathrm{src}}$, $C_{\mathrm{bkg}}$), effective exposure time ($t_{\mathrm{src}}$, $t_{\mathrm{bkg}}$), and apertures areas in units of pixels$^2$ ($A_{\mathrm{src}}$, $A_{\mathrm{bkg}}$) for both the source and background regions for each galaxy using the CIAO tool \texttt{dmextract}.  The effective exposure accounts for variations across the field of view due to the telescope optics, CCD gaps, and bad pixels.  The background region counts were extracted from the annular background regions, from which any X-ray detected sources as well as all MOSDEF galaxies were masked out.  Our measurements of the aperture areas account for any fraction of the area that was masked out. \par
For each source, we calculated the net background-subtracted counts and a conversion factor to translate the net counts into the rest-frame X-ray luminosity.  The net source counts ($C_{\mathrm{net}}$) are calculated as:
\begin{equation}
C_{\mathrm{net}} = C_{\mathrm{src}} - C_{\mathrm{bkg}}\times\frac{t_{\mathrm{src}}A_{\mathrm{src}}}{t_{\mathrm{bkg}}A_{\mathrm{bkg}}}
\end{equation}
Converting the net counts in the 0.5--2~keV band into rest-frame 2--10~keV luminosities requires assuming a spectral model to calculate the mean energy per photon ($E_{\mathrm{avg}}$) and the $k$-correction ($k_{\mathrm{corr}}$).  We assume an unobscured power-law spectrum with $\Gamma=2.0$ to facilitate comparison with the local $L_{\mathrm{X}}$-SFR-$Z$ relation measured by \citet{brorby16}.  In Section \ref{sec:systematics}, we discuss the validity and uncertainty associated with this assumption.   For each source, we calculate the conversion factor, $\omega$, between net counts and X-ray luminosity, as given by:
\begin{equation}
\begin{split}
L_{\mathrm{X}} & = C_{\mathrm{net}}/\omega \\
 \omega & = (t_{\mathrm{src}}\mathrm{ECF})/(4\pi D_{\mathrm{L}}^{2} E_{\mathrm{avg}} k_{\mathrm{corr}})
\end{split}
\end{equation}
where $D_{\mathrm{L}}$ is the luminosity distance and ECF $=0.9$ since our aperture regions are based on 90\% PSF radius.  \par

\subsection{Stacked X-ray luminosities}
For a given galaxy stack, we summed the net counts and the expected number of background counts from individual source apertures.  The average X-ray luminosity ($\langle L_{\mathrm{X}} \rangle$) of a stack of $N$ galaxies is calculated as the total net counts divided by the sum of the conversion factors:
\begin{equation}
\langle L_{\mathrm{X}} \rangle = \frac{1}{N}\sum_i^N \frac{C_{\mathrm{net,}i}}{\omega_{i}} \approx \frac{\sum_i^N C_{\mathrm{net,}i}}{\sum_i^N \omega_{i}}
\end{equation}
 This approximation is appropriate when the galaxies in a given stack have similar $L_X$.  We expect the range of $L_X$ for galaxies in a given $Z$ bin to primarily be set by the range of SFR, and therefore to span at most 2 orders of magnitude.  For such a range of $L_X$, we estimate that the approximation we use to measure $\langle L_{\mathrm{X}} \rangle$ should be accurate to 0.1 dex.   The conversion factor $\omega$ associated with each galaxy provides a relative weighting for how much each galaxy contributes to the measured X-ray emission.  Since our goal is to study the relationship between $L_{\mathrm{X}}$ and different galaxy properties (i.e. SFR, $Z$, $z$, $M_*$, sSFR), for each galaxy stack we also calculate the weighted average of each galaxy property, applying the $\omega$ factors used to calculate the X-ray luminosity.  We estimate that our measurement of $\langle L_{\mathrm{X}} \rangle/\langle $SFR$ \rangle$ should approximate $\langle L_{\mathrm{X}}/$SFR$ \rangle$ with an accuracy of 0.05 dex based on the scatter of 0.3-0.4 dex observed in the local $L_X$-SFR-$Z$ relation \citep{brorby16}.   \par
We computed two sources of error on each stacked signal.  The first is Poisson noise associated with the background, which we used to establish the significance of the signal in each stack.  We calculated the Poisson probability that a random fluctuation of the estimated background counts could result in a number of counts within the source regions greater than or equal to the total stacked counts (source plus background). \par
Table \ref{tab:prop} provides information about the properties of galaxies in our stacks, including the weighted average redshift, $M_*$, and SFR, as well as the median stacked O/H.  The X-ray properties of our stacks, including the signal significance, net counts, and mean $L_{\mathrm{X}}$, are presented in Table \ref{tab:xray}.
We tested that our stacking procedure does not result in an unusual number of spurious detections by stacking the \textit{Chandra} data at random sky positions rather than galaxy positions.  We apply the same X-ray selection criteria to the random positions as our galaxy sample.
We make 500 mock stacks for each of the 10 real stacks described in Table \ref{tab:prop}; each mock stack includes the same number of individual positions per \textit{Chandra} survey field as its corresponding real stack.  For each mock stack, we calculate the probability that the total stacked counts could be due to a random fluctuation of the estimated background.  The resulting probability distributions of the mock stacks is consistent with expectations for random noise (e.g. $1\%$ of mock stacks have a 1\% probability of being due to random noise).  Thus, the detection probabilities of our real stacks are reliable.
\par
The statistical uncertainties associated with the measured X-ray luminosities are calculated using a bootstrapping method, which measures how the contribution of individual sources affects the average stacked signal.  To determine the bootstrapping errors, we randomly resampled the galaxies in each stack 1000 times and repeated our stacking analysis.  The number of galaxies in a given stack is conserved during the resampling, leading some values to be duplicated while others are eliminated in a particular iteration.  From the resulting distribution of stacked X-ray luminosities, we measure 1$\sigma$ confidence intervals for stacked signals exceeding our detection threshold. 
\par
As done in \citet{lehmer16}, the uncertainties associated with the weighted average galaxy properties ($\langle Q_{\mathrm{phys}} \rangle$) are also calculated by a bootstrapping technique.  Each galaxy value is perturbed according to its error distribution and the weighted average is recalculated.  The calculation of these perturbed average values ($\langle Q^{\mathrm{pert}}_{\mathrm{phys,}k} \rangle$) is performed 1000 times and the 1$\sigma$ uncertainty on the weighted average is according to the following equation:
\begin{equation}
\sigma_{\langle Q_{\mathrm{phys}} \rangle} = \frac{1}{N_{\mathrm{boot}}}\Big[\sum_{\mathrm{k}=0}^{N_{\mathrm{boot}}}(\langle Q^{\mathrm{pert}}_{\mathrm{phys,}k} \rangle - \langle Q_{\mathrm{phys}} \rangle )^2 \Big]^{1/2}
\end{equation}
\par
The weighted average galaxy properties for each stack discussed in \S\ref{sec:results} are provided in Table \ref{tab:prop}, while the stacked X-ray properties are listed in Table \ref{tab:xray}.

\begin{table*}
\begin{minipage}{\textwidth}
\centering
\footnotesize
\caption{Average Galaxy Properties of Stacks}
\begin{tabular}{cccccccclcc} \hline \hline
\T \multirow{2}{*}{Stack ID} & \multicolumn{4}{c}{\# Galaxies} & \multirow{2}{*}{$\langle z \rangle$} & log$\langle M_* \rangle$ & \multirow{2}{*}{12+log(O/H)} & \hspace{0.3cm}SFR & $\langle \mathrm{SFR}\rangle$ & log$\langle \mathrm{sSFR} \rangle$ \\  \cline{2-5}
& All & AEGIS & CDF-N & CDF-S & & ($M_{\odot}$) & & Indicator & ($M_{\odot}$ yr$^{-1}$) & (yr$^{-1}$) \\
\B (a) & (b) & (c) & (d) & (e) & (f) & (g) & (h) & (i) & (j) & (k) \\
\hline
\multicolumn{5}{l}{\textbf{All sSFR}} \\
\hline
\T \B 1 & 79 & 53 & 22 & 3 & 1.92 & $10.121^{+0.002}_{-0.001}$ & $8.31\pm0.01$ & H$\alpha$, corr & $22.9^{+0.7}_{-0.3}$ & $-8.738^{+0.009}_{-0.004}$\\
\hline
\multicolumn{5}{l}{\textbf{All sSFR: redshift binning}} \\
\hline
\T 2 & 20 & 13 & 7 & 0 & 1.51 & $10.053^{+0.002}_{-0.001}$ & $8.32\pm0.02$ & H$\alpha$, corr & $16.6^{+0.5}_{-0.2}$ & $-8.856^{+0.010}_{-0.004}$\\
\B 3 & 59 & 41 & 15 & 3 & 2.26 & $10.172\pm0.001$ & $8.31\pm0.01$ & H$\alpha$, corr & $28.3^{+0.9}_{-0.4}$ & $-8.658^{+0.009}_{-0.004}$\\
\hline
\multicolumn{5}{l}{\textbf{All sSFR: metallicity binning}} \\
\hline
\T 4 & 30 & 19 & 8 & 3 & 2.05 & $9.971^{+0.002}_{-0.001}$ & $8.23^{+0.01}_{-0.02}$ & H$\alpha$, corr & $22.6^{+0.7}_{-0.3}$ & $-8.606^{+0.006}_{-0.003}$ \\
5 & 23 & 19 & 4 & 0 & 1.82 & $10.110^{+0.002}_{-0.001}$ & $8.35\pm0.01$ & H$\alpha$, corr & $27.3^{+0.8}_{-0.3}$ & $-8.711^{+0.013}_{-0.005}$ \\
\B 6 & 19 & 13 & 6 & 0 & 1.78 & $10.337\pm0.002$ & $8.52\pm0.02$ & H$\alpha$, corr & $25.4^{+0.9}_{-0.3}$ & $-8.983^{+0.014}_{-0.005}$ \\
\hline
\multicolumn{5}{l}{\textbf{Restricting sSFR: metallicity binning}} \\
\hline
\multicolumn{5}{l}{\textbf{\hspace{0.1cm}$-9.3<$ log(sSFR)$_{\mathrm{H}\alpha, \mathrm{corr}}$$<-8.4$}} \\ \cline{1-4}
7a & 24 & 15 & 7 & 2 & 2.04 & $10.001\pm0.001$ & $8.24^{+0.01}_{-0.02}$ & H$\alpha$, corr & $20.6^{+0.6}_{-0.2}$ & $-8.705^{+0.005}_{-0.002}$ \\
8a & 21 & 17 & 4 & 0 & 1.83 & $10.110\pm0.001$ & $8.35\pm0.01$ & H$\alpha$, corr & $23.4^{+0.7}_{-0.3}$ & $-8.791^{+0.013}_{-0.005}$ \\
9a & 15 & 11 & 4 & 0 & 1.86 & $10.408^{+0.002}_{-0.001}$ & $8.45^{+0.02}_{-0.03}$ & H$\alpha$, corr & $36.2^{+1.3}_{-0.5}$ & $-8.849^{+0.014}_{-0.006}$ \\
\cline{1-4}
\multicolumn{5}{l}{\textbf{\T \hspace{0.1cm}$-9.2<$ log(sSFR)$_{\mathrm{H}\alpha}$ $<-8.4$}} \\ \cline{1-4}
7b & 18 & 9 & 7 & 1 & 2.01 & $10.019\pm0.001$ & $8.22\pm0.02$ & H$\alpha$ & $26.0^{+0.7}_{-0.3}$ & $-8.650^{+0.011}_{-0.004}$ \\
8b & 20 & 17 & 3 & 0 & 1.94 & $10.177\pm0.001$ & $8.38\pm0.01$ & H$\alpha$ & $28.9^{+0.9}_{-0.4}$ & $-8.720^{+0.014}_{-0.005}$ \\
9b & 14 & 10 & 4 & 0 & 1.89 & $10.343^{+0.002}_{-0.001}$ & $8.45^{+0.02}_{-0.03}$ & H$\alpha$ & $36.0^{+1.3}_{-0.5}$ & $-8.827^{+0.014}_{-0.006}$ \\
\cline{1-4}
\multicolumn{5}{l}{\textbf{\T \hspace{0.1cm}$-9.05<$ log(sSFR)$_{\mathrm{SED}}$ $<-8.5$}} \\ \cline{1-4}
7c & 22 & 13 & 8 & 1 & 2.05 & $9.980\pm0.001$ & $8.25^{+0.01}_{-0.02}$ & SED & $19.4^{+0.2}_{-0.1}$ & $-8.663^{+0.005}_{-0.003}$ \\
8c & 22 & 18 & 4 & 0 & 1.81 & $10.113^{+0.002}_{-0.001}$ & $8.35^{+0.01}_{-0.01}$ & SED & $22.0\pm0.2$ & $-8.801^{+0.005}_{-0.003}$ \\
\B9c & 15 & 10 & 5 & 0 & 1.75 & $10.192^{+0.002}_{-0.001}$ & $8.50\pm0.02$ & SED & $25.3^{+0.3}_{-0.2}$ & $-8.851^{+0.005}_{-0.003}$ \\
\hline
\multicolumn{5}{l}{\textbf{High sSFR: log(sSFR) $>-8.8$}} \\
\hline
\T \B 10 & 37 & 26 & 6 & 3 & 2.03 & $10.117^{+0.002}_{-0.001}$ & $8.30^{+0.01}_{-0.02}$ & H$\alpha$, corr & $37.5^{+1.2}_{-0.5}$ & $-8.508^{+0.009}_{-0.004}$ \\
\hline
\multicolumn{5}{l}{\textbf{High sSFR: redshift binning}} \\
\hline
\T 11 & 8 & 7 & 1 & 0 & 1.49 & $10.140^{+0.003}_{-0.001}$ & $8.31^{+0.01}_{-0.03}$ & H$\alpha$, corr & $39.0^{+1.2}_{-0.5}$ & $-8.510^{+0.011}_{-0.005}$ \\
\B 12 & 29 & 21 & 5 & 3 & 2.27 & $10.106\pm0.001$ & $8.29^{+0.01}_{-0.02}$ & H$\alpha$, corr & $36.9^{+1.2}_{-0.5}$ & $-8.508^{+0.008}_{-0.004}$\\
\hline
\multicolumn{5}{l}{\textbf{High sSFR: metallicity binning}} \\
\hline
\T 13 & 19 & 12 & 4 & 3 & 2.12 & $10.012^{+0.002}_{-0.001}$ & $8.23\pm0.02$ &  H$\alpha$, corr & $30.2^{+0.9}_{-0.4}$ & $-8.493^{+0.006}_{-0.003}$ \\
\B 14 & 17 & 15 & 2 & 0 & 1.87 & $10.264^{+0.002}_{-0.001}$ & $8.40\pm0.01$ &  H$\alpha$, corr & $51.6^{1.7}_{-0.7}$ & $-8.530^{+0.014}_{-0.006}$ \\
\hline
\hline \hline
\multicolumn{11}{p{7.0in}}{\T Notes:

(h) X-ray weighted, median oxygen abundance of the galaxy stack based on O3N2 indicator \citep{pettini04}.  

(i) SED SFR listed assumes $Z=0.02$, Calzetti extinction curve, and constant SF history.  The H$\alpha$ indicator assumed $Z=0.02$ and the Cardelli extinction curve.  For the H$\alpha$, corr indicator, the conversion factor for galaxies with 12+log(O/H)$<8.3$ assumes $Z=0.004$ and the Cardelli extinction curve.
} \\
\end{tabular}
\label{tab:prop}
\end{minipage}
\end{table*}

\begin{table*}
\begin{minipage}{\textwidth}
\centering
\footnotesize
\caption{Stacked X-ray Properties}

\begin{tabular}{cccccclc} \hline \hline
\T \multirow{2}{*}{Stack ID} & Total exposure & Effective exposure & \multirow{2}{*}{$P_{\mathrm{random}}$} & \multirow{2}{*}{Net counts} & $\langle L_{\mathrm{X}} \rangle$ & SFR & \multirow{2}{*}{log$\frac{\langle L_{\mathrm{X}}\rangle}{\langle \mathrm{SFR}\rangle_{H\alpha,\mathrm{corr}}}$}\\
 & (Ms) & (Ms cm$^{-2}$) & & & ($10^{40}$ erg s$^{-1}$) & Indicator & \\
 (a) & (b) & (c) & (d) & (e) & (f) & (g) & (h)\\
 \hline
\multicolumn{5}{l}{\textbf{All sSFR}} \\
\hline
\T 1 & 105.1 & 25779 & 1.5e-8 & $96\pm19$ & $19.9^{+3.1}_{-4.4}$ & H$\alpha$, corr & $39.94^{+0.06}_{-0.11}$\\
\hline
\multicolumn{5}{l}{\textbf{All sSFR: redshift binning}} \\
\hline
\T 2 & 23.8 & 6204 & 3.7e-4 & $26^{+10}_{-9}$ & $11.8^{+2.2}_{-4.3}$ & H$\alpha$, corr & $39.85^{+0.08}_{-0.19}$\\
\B 3 & 81.3 & 19575 & 3.5e-6 & $70\pm17$ & $26.8^{+5.6}_{-7.1}$ & H$\alpha$, corr & $39.98^{+0.08}_{-0.13}$ \\
 \hline
\multicolumn{5}{l}{\textbf{All sSFR: metallicity binning}} \\
\hline
\T 4 & 50.5 & 12029 & 3.8e-4 & $40\pm13$ & $20.5^{+4.3}_{-6.8}$ & H$\alpha$, corr & $39.96^{+0.08}_{-0.18}$ \\
 5 & 22.6 & 5422 & 2.3e-4 & $27^{+10}_{-9}$ & $24.4^{+11.7}_{-4.9}$ & H$\alpha$, corr & $39.95^{+0.17}_{-0.10}$ \\
\B 6 & 21.8 & 5576 & 3.9e-3 & $22^{+10}_{-9}$ & $17.5^{+5.6}_{-7.7}$ & H$\alpha$, corr & $39.84^{+0.12}_{-0.25}$ \\
\hline
\multicolumn{5}{l}{\textbf{Restricted sSFR: metallicity binning}} \\
\hline
\T 7a & 38.7 & 9245 & 5.8e-6 & $42\pm11$ & $27.4^{+7.0}_{-5.2}$ & H$\alpha$, corr & $40.12^{+0.10}_{-0.09}$ \\
8a & 21.1 & 5046 & 4.6e-4 & $25^{+9}_{-8}$ & $23.8^{+13.9}_{-6.2}$ & H$\alpha$, corr & $40.01^{+0.20}_{-0.13}$ \\
9a & 16.4 & 4080 & 9.2e-3 & $18^{+9}_{-8}$ & $21.4^{+8.8}_{-11.2}$ & H$\alpha$, corr & $39.77^{+0.15}_{-0.32}$\\
\\
7b & 27.3 & 6868 & 8.2e-6 & $32^{+10}_{-9}$ & $27.6^{+7.7}_{-5.6}$ & H$\alpha$ & $40.03^{+0.11}_{-0.10}$ \\
8b & 19.1 & 4513 & 1.6e-4 & $26^{+9}_{-8}$ & $32.1^{+8.8}_{-9.5}$ & H$\alpha$ & $40.04^{+0.11}_{-0.15}$ \\
9b & 15.6 & 3880 & 1.4e-2 & $16^{+9}_{-8}$ & $21.5^{+9.4}_{-11.9}$ & H$\alpha$ & $39.78^{+0.16}_{-0.35}$\\
\\
7c & 32.4 & 8004 & 5.0e-4 & $30\pm10$ & $23.1^{+8.0}_{-6.5}$ & SED & $40.08^{+0.13}_{-0.14}$ \\
8c & 21.8 & 5215 & 5.6e-4 & $24^{+9}_{-8}$ & $22.0^{+12.2}_{-4.4}$ & SED & $40.00^{+0.19}_{-0.10}$\\
\B 9c & 17.5 & 4587 & 2.6e-2 & $14^{+9}_{-8}$ & $13.1^{+1.5}_{-4.4}$ & SED & $39.71^{+0.05}_{-0.18}$\\
\hline
\multicolumn{5}{l}{\textbf{High sSFR: log(sSFR) $>-8.8$}} \\
\hline
 \T 10 & 52.1 & 12656 & 2.4e-5 & $54\pm14$ & $25.8^{+5.8}_{-5.5}$ & H$\alpha$, corr & $39.84^{+0.09}_{-0.11}$ \\
  \hline
\multicolumn{5}{l}{\textbf{High sSFR: redshift binning}} \\
\hline
\T 11 & 7.4 & 1717 & 8.1e-3 & $11^{+6}_{-5}$ & $16.7^{+4.6}_{-7.6}$ & H$\alpha$, corr & $39.63^{+0.11}_{-0.26}$ \\
\B 12 & 46.2 & 10939 & 2.9e-4 & $43\pm14$ & $29.8^{+8.0}_{-7.7}$ & H$\alpha$, corr & $39.91^{+0.10}_{-0.13}$ \\
 \hline
\multicolumn{5}{l}{\textbf{High sSFR: metallicity binning}} \\
\hline
\T 13 & 37.3 & 8738 & 2.8e-3 & $30\pm12$ & $23.1^{+7.9}_{-6.2}$ & H$\alpha$, corr & $39.88\pm0.13$\\
 \B 14 & 15.6 & 3731 & 3.5e-3 & $19^{+9}_{-8}$ & $26.4^{+6.4}_{-8.0}$ & H$\alpha$, corr & $39.71^{+0.09}_{-0.15}$\\
  \hline \hline
 \multicolumn{8}{p{6.0in}}{\T Notes:
 
 (c) Total exposure multiplied by \textit{Chandra} effective area. 
 
 (e) Errors are based on Poisson statistics.
 
 (g) Errors are based on bootstrapping.
 } \\
\end{tabular}
\label{tab:xray}
\end{minipage}
\end{table*}

\subsection{Stacked Metallicity Measurements}
To maximize our sample size and reduce biases in our galaxy sample, in our stacking analysis we include galaxies with upper or lower limits on their oxygen abundance based on the O3N2 indicator.  For parts of our analysis we split our sample into different $Z$ bins.  In the highest $Z$ bin, we include galaxies with 12+log(O/H) lower limits higher than the bin's lower bound, and in the lowest $Z$ bin, we include galaxies with 12+log(O/H) upper limits lower than the bin's upper bound.  Overall, for the 79 galaxies in our full sample, we have 59 O/H measurements, 18 upper limits, and 2 lower limits. \par
Due to the inclusion of upper and lower limits on 12+log(O/H), we cannot simply average the O/H values of the galaxies in each stack.  Instead, using the method of \citet{sanders15}, we measure the stacked O/H by making composite spectra of the galaxies in each stack.  Each galaxy spectrum was shifted to rest frame, converted from flux density to luminosity density, corrected for reddening assuming the \citet{cardelli89} attenuation curve, interpolated onto a common wavelength grid, and normalized by the H$\alpha$ luminosity.
Normalized composite spectra were created by taking the X-ray weighted ($\omega$) median value of the normalized spectra.  Emission-line luminosities were measured from the composite spectra by fitting a flat continuum and Gaussian profiles to regions around emission features.  Uncertainties were estimated using a Monte Carlo technique.  Testing this method using only galaxies with detections of all lines of interest, this stacking method was found to robustly reproduce the X-ray weighted median line ratios of the galaxies in a stack.  \par

\section{Results and Discussion}
\label{sec:results}

\subsection{The redshift evolution of XRBs}
\label{sec:zevol}
\begin{figure}
\centering
\includegraphics[width=0.5\textwidth]{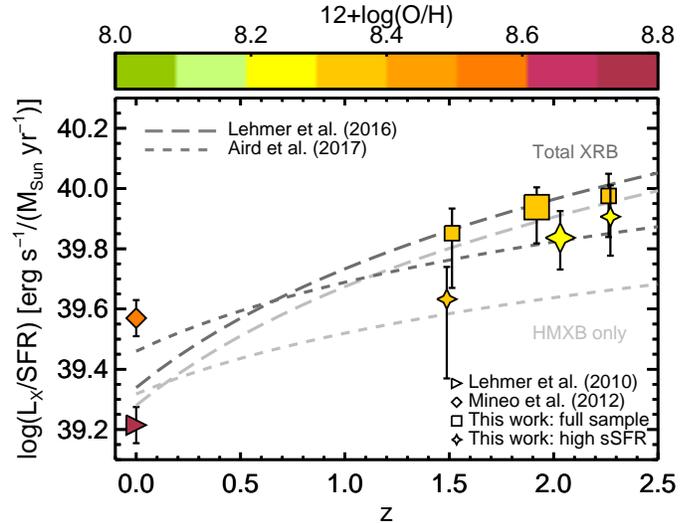}
\vspace{0.25in}
\caption{Stacked $L_{\mathrm{X}}$/SFR values of our galaxy sample split into $z\sim1.5$ and $z\sim2.3$ redshift bins are shown by small squares/stars; squares represent stacks of the full galaxy sample, stars represent stacks of high sSFR (sSFR$>10^{-8.8}$ yr$^{-1}$) galaxies that are HMXB-dominated, and colors represent the weighted median oxygen abundance of the stacks.  The larger square (star) symbol represents the combined $z\sim2$ stack of all (high sSFR) galaxies.  The colored diamond and triangle represent local ($z=0$) measurements of the $L_{\mathrm{X}}$-SFR relation.
The long and short dashed lines show the redshift evolution of $L_{\mathrm{X}}$/SFR for the total XRB (dark gray) and HMXB-only (light gray) emission derived by \citet{lehmer16} and \citet{aird17}; since  the \citetalias{lehmer16} total XRB evolution and the \citetalias{aird17} HMXB-only evolution are parametrized as non-linear relations between $L_{\mathrm{X}}$ and SFR, these curves have been normalized for SFR$=20 M_{\odot}$ yr$^{-1}$, the mean SFR of our $z\sim2$ galaxy sample.  Our stacks lie above the $z=0$ measurements and are consistent with the redshift evolution measured by previous studies. }
\label{fig:lxzdep}
\end{figure}

We first investigate whether our galaxy sample supports the redshift evolution of $L_{\mathrm{X}}$/SFR of XRBs found by previous studies (e.g. \citealt{lehmer16}; \citealt{aird17}).  We use X-ray stacks of our full $z\sim2$ sample as well as the subsample of galaxies with sSFR$>10^{-8.8}$ yr$^{-1}$.  These high sSFR stacks should be dominated by HMXBs, while the full sample stacks likely contain significant contributions from both LMXBs and HMXBs, as discussed in more detail in \S\ref{sec:hmxb}.  Since our galaxy sample is bimodally distributed in redshift due to atmospheric windows (see Figure \ref{fig:redshift}), we also split the galaxy sample between these two redshift intervals: $1.3<z<1.7$ and $2.0<z<2.6$.  Information for all the aforementioned stacks is provided in rows \#$1-3$ and $10-12$ of Tables \ref{tab:prop} and \ref{tab:xray}.  The full sample redshift stacks are represented by colored squares in Figure \ref{fig:lxzdep}, while the high sSFR stacks are shown by colored stars.  \par
In Figure \ref{fig:lxzdep}, we also show $L_{\mathrm{X}}$/SFR values measured by two studies (\citealt{lehmer10}; \citealt{mineo12}, hereafter \citetalias{lehmer10} and \citetalias{mineo12}, respectively) using samples of nearby galaxies at $z=0$.  The \citetalias{mineo12} value is converted from the $0.5-8$ to $2-10$ keV band assuming $\Gamma=2.0$ and $N_{\mathrm{H}}=3\times10^{21}$ cm$^{-2}$, the average column density measured for their galaxy sample.  Both the \citetalias{lehmer10} and \citetalias{mineo12} values are converted to be consistent with a Chabrier IMF.  The 0.3 dex difference between these local values is likely due to sample selection effects.  In fact, it is possible that the average metallicity of the galaxies in the \citetalias{lehmer10} and \citetalias{mineo12} differs significantly.  Metallicity information is not available for all the galaxies in the \citetalias{lehmer10} or \citetalias{mineo12} samples.  Therefore, we estimate the mean O/H for the \citetalias{lehmer10} sample using the $M_*-Z$ relation from \citet{kewley08} based on the O3N2 calibration by \citep{pettini04}, and by combining the O/H measurements for 19 of the \citetalias{mineo12} galaxies gathered by \citet{douna15} and estimates based on the $M_*-Z$ relation for the remaining 10 \citetalias{mineo12} galaxies.  As shown by the colorbar in Figure \ref{fig:lxzdep}, we find that the mean O/H of the \citetalias{lehmer10} sample is much higher (12+log(O/H) $=8.71$) than the value for the \citetalias{mineo12} sample (12+log(O/H) $=8.57$).  This difference could contribute to the discrepancy between these two local measurements of $L_{\mathrm{X}}$/SFR.
\par
 Both the $z\sim1.5$ and $z\sim2.3$ stacks lie above the local $L_{\mathrm{X}}$/SFR values, although due to the large error bars, the $z\sim1.5$ stacks are statistically consistent with the \citetalias{mineo12} local value.  The $L_{\mathrm{X}}$/SFR of the $z\sim2.3$ stacks are higher than, but statistically consistent with, the $z\sim1.5$ stacks.  The $L_{\mathrm{X}}$/SFR of the $z\sim2.3$ full sample stack is 2.3$\sigma$ higher than the \citetalias{mineo12} value of $3.7\times10^{39}$ erg s$^{-1}$ $M_{\odot}$ yr and $3.1\sigma$ higher than the \citetalias{lehmer10} value of $1.6\times10^{39}$ erg s$^{-1}$ $M_{\odot}$ yr.  \citet{fornasini18} find that for X-ray stacks with $\lesssim50$ galaxies such as these, $\langle L_{\mathrm{X}} \rangle$ may be biased to higher values than the true mean by 0.15 dex;\footnote{For small galaxy sample, $\langle L_{\mathrm{X}} \rangle$ can be biased to high values due to insufficient sampling of the XRB luminosity function.} however, even accounting for this possible systematic effect, the $z\sim2.3$ stacks have enhanced $L_{\mathrm{X}}$/SFR compared to the \citetalias{lehmer10} and \citetalias{mineo12} relations.  Both the $z\sim1.5$ and $z\sim2.3$ full sample stacks are in good agreement with the $L_{\mathrm{X}}$/SFR values expected from the redshift evolution of the total X-ray binary (XRB) emission measured by \citet{lehmer16} (shown by the dark gray long-dashed line in Figure \ref{fig:lxzdep}; hereafter \citetalias{lehmer16}) and \citet{aird17} (shown by the dark gray short-dashed line; hereafter \citetalias{aird17}).  \par
Both these previous studies also decompose the total XRB emission into an LMXB contribution proportional to $M_*$ and an HMXB contribution proportional to SFR; the latter is traced by the light gray lines in Figure \ref{fig:lxzdep}.  The high sSFR stacks, which represent the most HMXB-dominated galaxies, are consistent (within 1.7$\sigma$) with the HMXB-only redshift evolution measured by \citetalias{lehmer16} and \citetalias{aird17}.  Thus, our galaxy sample supports the redshift evolution of $L_{\mathrm{X}}$/SFR measured by other works.  

\subsection{The metallicity dependence of XRBs}
\label{sec:zdep}
Having established that our $z\sim2$ galaxy sample does show enhanced $L_{\mathrm{X}}$/SFR relative to $z=0$ galaxies, we investigate whether this enhancement could be driven by the $Z$ dependence of HMXBs.  We tried simultaneously splitting our sample by redshift and $Z$ but we do not have a sufficiently high signal-to-noise ratio to obtain meaningful results.  Therefore, we instead split our full sample with $1.3<z<2.6$ by $Z$, and find that splitting the sample into three bins with divisions at 12+log(O/H) $=8.3$ and 8.4 yields detections with $>2.5\sigma$ significance (see stacks \# $4-6$ in Tables \ref{tab:prop} and \ref{tab:xray}).  \par

\begin{figure*}
\centering
\includegraphics[width=0.85\textwidth]{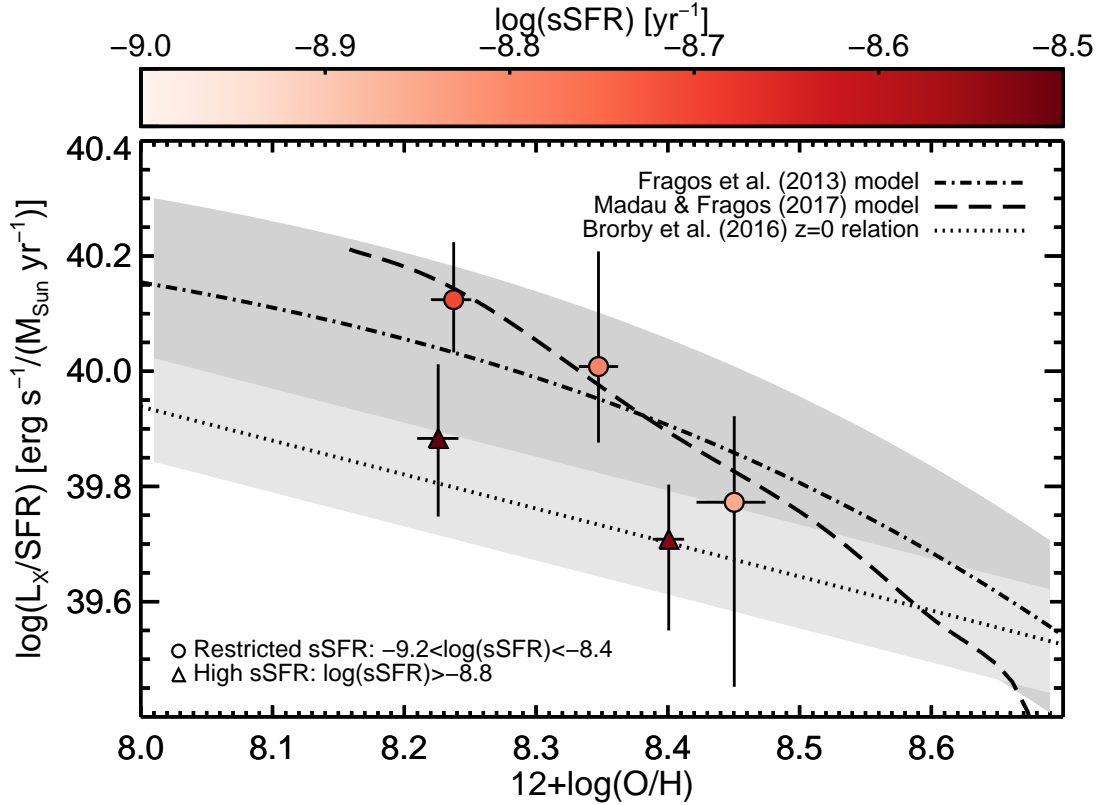}
\vspace{0.5in}
\caption{$L_{\mathrm{X}}$/SFR versus oxygen abundance measurements for stacks of galaxies selected using different sSFR criteria, colored according to sSFR.  The dotted line shows the local $L_{\mathrm{X}}$-SFR-$Z$ relation from \citetalias{brorby16} based on nearby galaxies spanning $7.0<$ 12+log(O/H) $<9.0$ with corresponding error shown in light gray.  The mean of the six best-fitting models from \citetalias{fragos13b} is shown as a dash-dotted line, with the parameter space covered by these six models shown in dark gray.  The best-fit model from \citetalias{madau17} is shown by a dashed line, and has been converted from the \citet{kk04} $R_{23}$ scale to the O3N2 scale from \citet{pettini04} using the conversion from \citet{kewley08}.  Results for a restricted sSFR range that yields a similar sSFR distribution across all $Z$ are shown by circles; the $L_{\mathrm{X}}$/SFR of these stacks favors a $Z$-dependent model.  Results for high sSFR stacks are shown by triangles; this sample provides the cleanest measurement of HMXB-only $L_{\mathrm{X}}$/SFR and is consistent with the local $L_{\mathrm{X}}$-SFR-$Z$ relation and some theoretical models.}
\label{fig:lxssfr}
\end{figure*}

These three $Z$ stacks show a slight hint of an anti-correlation between $L_{\mathrm{X}}$/SFR and O/H. 
However, given the large statistical errors on $L_{\mathrm{X}}$, the $L_{\mathrm{X}}$/SFR values of all these bins are consistent within 1$\sigma$ with a constant ($Z$-independent) value.  Thus, from these stacks, it is not possible to determine whether the redshift evolution of $L_{\mathrm{X}}$/SFR is driven by metallicity or some other factor since both a $Z$-dependent and $Z$-independent model can fit the data.   \par 
However, it is important to consider if systematic factors may impact this result.  As mentioned in \S \ref{sec:sample}, a key variable which is known to affect $L_{\mathrm{X}}$/SFR is the sSFR \citep{lehmer10}.  Since $L_{\mathrm{HMXB}}$ is correlated with SFR and $L_{\mathrm{LMXB}}$ is correlated with $M_*$, galaxies with lower sSFR have higher $L_{\mathrm{X}}$/SFR due to the larger fractional contribution of LMXBs to the X-ray emission.  \par
The average sSFR and $Z$ of the three stacks based our full sample are strongly anti-correlated.  As reported in Table \ref{tab:prop}, the average sSFR of these three stacks spans 0.4 dex, and $L_{\mathrm{X}}$/SFR can vary by up to 0.3 dex over this sSFR range \citep{lehmer10}.  Such variation is comparable in magnitude to the predicted decrease of $L_{\mathrm{X}}$/SFR with $Z$ for the $Z$ range probed by our stacks (\citealt{fragos13b}; \citealt{madau17}; hereafter \citetalias{fragos13b} and \citetalias{madau17}, respectively).  Thus, $L _{\mathrm{X}}$/SFR could be inflated for the highest $Z$ stack due to its lower sSFR and deflated for the lowest $Z$ stack due to its higher sSFR.  As a result, the fact that $Z$ and sSFR are anti-correlated could artificially mask the expected decrease of $L_{\mathrm{X}}$/SFR with $Z$.  \par
Therefore, to reduce possible systematic effects associated with sSFR, we further restricted our sample to galaxies with $-9.3<$ log(sSFR) $<-8.4$, a range of sSFR values common across all $Z$.  This sSFR-matching criterion reduces the spread of weighted average sSFR values of the stacks to 0.15 dex (see stacks \# $7a-9a$ in Tables \ref{tab:prop} and \ref{tab:xray}).  This residual anti-correlation between $Z$ and sSFR that exists in the sSFR-restricted stacks may still slightly flatten the intrinsic $L_{\mathrm{X}}$-SFR-$Z$ relation, but the impact on the stacked $L_{\mathrm{X}}$/SFR should be $<0.1$ dex.  \par
The resulting $L_{\mathrm{X}}$/SFR values versus $Z$ are shown by the circles in Figure \ref{fig:lxssfr}.
The sSFR-restricted stacks are not consistent within 1$\sigma$ uncertainties with a constant ($Z$-independent) $L_{\mathrm{X}}$/SFR value.  We estimate the significance of the observed anti-correlation by calculating the probability that the data points are described by a power-law relation between $L_{\mathrm{X}}$/SFR and $Z$ with a negative index rather than an index $\geq$0.  The probability that the data are consistent with a negative correlation between $L_{\mathrm{X}}$/SFR and $Z$ is 97\%.  This result is robust to the statistical uncertainties in the stacked O/H values.  Thus, this study provides the first direct evidence that the $L_{\mathrm{X}}$/SFR of XRBs at $z>0$ is anti-correlated with $Z$, although due to the low X-ray statistics, this conclusion is only supported with $\approx2\sigma$ confidence. \par
The  redshift of the sSFR-restricted stacks is roughly anti-correlated with O/H.  Therefore, one might worry that the $Z$ dependence we observe is caused by the redshift evolution of $L_{\mathrm{X}}$/SFR, rather than vice versa.  However, the weighted average redshifts of the stacks vary by $<$0.2.  As found by previous studies and confirmed by our redshift-binned stacks,  $L_{\mathrm{X}}$/SFR evolves too slowly with redshift for such a small $z$ difference to account for the 0.35 dex difference in $L_{\mathrm{X}}$/SFR between our lowest and highest $Z$ stacks.  Therefore, the $Z$ dependence we measure cannot be attributed to the small variation in redshift between our stacks. \par
The sSFR-restricted stacks are in good agreement with the best-fit theoretical models of \citetalias{fragos13b} and \citetalias{madau17}, which predict an anti-correlation between $L_{\mathrm{X}}$/SFR and $Z$.  However, it should be noted that these models represent the X-ray emission of HMXBs alone, while our sSFR-restricted stacks likely include a significant LMXB contribution, and therefore some tension exists between our observations and these best-fit models (see \S\ref{sec:hmxb} for more details).\par
Based on the sSFR-restricted stacks alone, we cannot determine whether the observed anti-correlation between $L_{\mathrm{X}}$/SFR and $Z$ at $z\sim2$ is driven by HMXBs, LMXBs, or both.  The X-ray luminosity of LMXB populations is expected to decrease with increasing $Z$ for the same reason as for HMXBs \citep{fragos13a}, but the oxygen abundance of H{\small II} regions is not an adequate proxy for the metallicity of LMXBs, which are associated with the old stellar population.  Furthermore, the dependence of $L_{\mathrm{LMXB}}$ on the age of the stellar population is predicted to be much stronger than its dependence on $Z$ (\citetalias{fragos13b}; \citetalias{madau17}).  Thus, $L_{\mathrm{LMXB}}$ is not expected to depend directly on gas-phase O/H, but if gas-phase O/H and mean stellar age are positively correlated at $z\sim2$, then the dependence of $L_{\mathrm{LMXB}}/M_*$ on age may contribute to the observed anti-correlation between $L_{\mathrm{X}}$/SFR and $Z$.  In order for LMXBs to account for the majority of the observed trend, $L_{\mathrm{LMXB}}/M_*$ must be a factor of $\approx$7 higher in the lowest O/H stack compared to the highest O/H stack.  According to the \citetalias{fragos13b} and \citetalias{madau17} LMXB models, such a large difference in $L_{\mathrm{LMXB}}/M_*$ could be explained by a mean stellar age difference of $\gtrsim2$ Gyr.  However, the age of the Universe at $z\sim2$ is only 3.3 Gyr.  Thus, it seems probable that HMXBs are responsible for at least a significant fraction of the observed anti-correlation between $L_{\mathrm{X}}$/SFR and $Z$.  In \S\ref{sec:hmxb}, this hypothesis is tested more directly.

\subsection{Comparing HMXB populations at $z=0$ and $z\sim2$}
\label{sec:hmxb}
In order to determine whether the redshift evolution of HMXBs is driven by the $Z$ dependence of $L_{\mathrm{X}}$/SFR, we further need to isolate the HMXB contribution to the observed XRB $Z$ dependence at $z\sim2$ and test whether it is the same as that measured in the local Universe.
In order to compare the normalization of the $L_{\mathrm{X}}$-SFR-$Z$ relation for HMXBs at $z\sim2$ and $z=0$, it is critical to minimize the absolute LMXB contribution by focusing on high sSFR galaxies.  Isolating the HMXB contribution is necessary because the local relation has been determined using only HMXB-dominated galaxies with sSFR $>10^{-10}$ yr$^{-1}$ \citep{brorby16}.  This high sSFR selection is also important for comparing our $z\sim2$ stacks to the \citetalias{fragos13b} and \citetalias{madau17} theoretical predictions that are based on the HMXB population alone.  \par
\begin{figure*}
\centering
\includegraphics[width=0.85\textwidth]{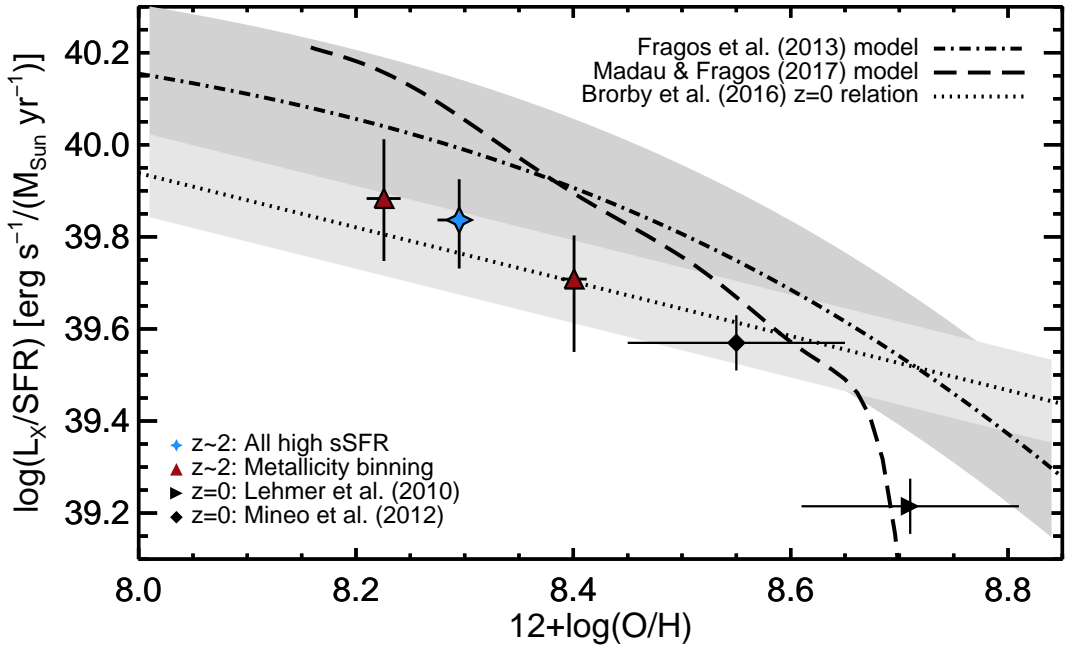}
\caption{$L_{\mathrm{X}}$/SFR versus stacked oxygen abundance for stacks of high sSFR (sSFR $>10^{-8.8}$) galaxies split by O/H are shown by triangles.  A stack of all sSFR galaxies at $z\sim2$ is shown by a four-point star.  These stacks provide the cleanest measurement of HMXB-only $L_{\mathrm{X}}$/SFR.  They are consistent with the local $L_{\mathrm{X}}$-SFR-$Z$ relation and the lower theoretical predictions from \citetalias{fragos13b}.  These stacks are inconsistent with the \citetalias{madau17} model and the upper bound of \citetalias{fragos13b} models at $3\sigma$ confidence.  The diamonds show local measurements of $L_{\mathrm{X}}$/SFR from \citet{lehmer10} and \citet{mineo12}; our estimates of the mean O/H values for these samples are described in \S\ref{sec:zevol}.  The lines shown in this figure are as described in the caption for Figure \ref{fig:lxssfr}.  }
\label{fig:lxhigh}
\end{figure*}

 
As discussed in \S\ref{sec:sample}, while all the MOSDEF galaxies in our sample have sSFRs higher than the sSFR $>10^{-10}$ yr$^{-1}$ threshold used to select HMXB-dominated galaxies in the local Universe \citep{lehmer10}, this transition value increases with redshift due to the evolution of $L_{\mathrm{X}}$/SFR of HMXBs and $L_{\mathrm{X}}/M_*$ of LMXBs \citep{lehmer16}.  Limiting our sample to log(sSFR) $>-8.8$, which is the approximate transition value found by \citet{lehmer16} at $z\sim2$, reduces the sample size by 50\%, resulting in very poor signal to noise in all but the lowest of the three $Z$ bins.  Therefore, we combine galaxies in the middle $Z$ and high $Z$ bins to obtain a statistically meaningful second measurement.  The $L_{\mathrm{X}}$/SFR values of the two high sSFR stacks are shown as triangles in Figures \ref{fig:lxssfr} and \ref{fig:lxhigh}, and the stack properties are provided in rows \# $13-14$ of Tables \ref{tab:prop} and \ref{tab:xray}.  The  $L_{\mathrm{X}}$/SFR values of the high sSFR stacks are lower than for the full sSFR and restricted sSFR samples, suggesting that there is a non-negligible LMXB contribution to the X-ray emission in the latter samples.  Based on the differences between stacks with similar $Z$ but different $\langle sSFR \rangle$, we estimate that $L_{\mathrm{X}}/M_*$ due to LMXBs is approximately $2-8\times10^{30}$ erg s$^{-1} M_{\odot}^{-1}$.  This $L_{\mathrm{X}}/M_*$ value is over an order of magnitude higher than local measurements for LMXB populations (\citealt{gilfanov04}; \citealt{colbert04}; \citealt{lehmer10}).  However, this value is consistent with the predictions of the six best-fitting population synthesis models of \citet{fragos13a} for $z\sim2$ and the best model of \citet{madau17} for LMXB populations with stellar mass-weighted ages of $1.5-3$~Gyr. \par
The two high sSFR stacks favor a negative correlation between $L_{\mathrm{X}}$/SFR and $Z$ with 86\% confidence.  While the significance of this result is not as high as for the restricted sSFR stacks, it suggests that the luminosity of HMXBs specifically, and not just XRBs generally, depends on $Z$ at $z\sim2$.  The high sSFR stacks are consistent with the local $L_{\mathrm{X}}$-SFR-$Z$ relation and the lower $L_{\mathrm{X}}$/SFR bound of the \citetalias{fragos13b} population synthesis models.  However, our high sSFR stacks exhibit significant ($>3\sigma$) tension with the upper $L_{\mathrm{X}}$/SFR bound of the \citetalias{fragos13b} models and the best-fit \citetalias{madau17} model.  Some of this tension may be due to remaining uncertainties in the absolute calibration of SFR indicators at $z\sim2$ and the absolute metallicity scale.
\par
Figure \ref{fig:lxhigh} shows the high sSFR stacks as well as the $z\sim2$ high sSFR stack and local $L_{\mathrm{X}}$/SFR measurements from \citetalias{lehmer10} and \citetalias{mineo12}.  We calculated the mean oxygen abundance of these local samples as described in \S\ref{sec:zevol}; since many of the oxygen abundance estimates are based on the $M_*-z$ relation, we show the scatter of the $M_*-z$ relation from \citet{kewley08} as horizontal error bars for these local points.  As shown, both the discrepancy between the local $L_{\mathrm{X}}$/SFR measurements and the enhanced $L_{\mathrm{X}}$/SFR of the $z\sim2$ stack can be explained by the anti-correlation of $L_{\mathrm{X}}$/SFR and $Z$.  If we combine the high sSFR $z\sim2$ stacks with at least one of the mean $L_{\mathrm{X}}$/SFR values measured for $z=0$, then an anti-correlation between $L_{\mathrm{X}}$/SFR and $Z$ is favored at $>99.7\% (>3\sigma)$  confidence.  The significance of this trend established by combining mean measurements at $z\sim2$ and $z=0$ is comparable to the significance of the trend measured by \citetalias{brorby16} using individually detected galaxies at $z=0$.\par
Thus, we find that the $Z$ dependence of the $L_{\mathrm{X}}$/SFR of HMXBs at $z\sim2$ is consistent with that measured for $z=0$ and some theoretical models.  This result provides the first direct link between the observed redshift evolution of $L_{\mathrm{X}}$/SFR and the $Z$ dependence of HMXBs.

\subsection{The impact of different SFR indicators}
\label{sec:sfrindicator}

\begin{figure*}
\centering
\includegraphics[width=0.85\textwidth]{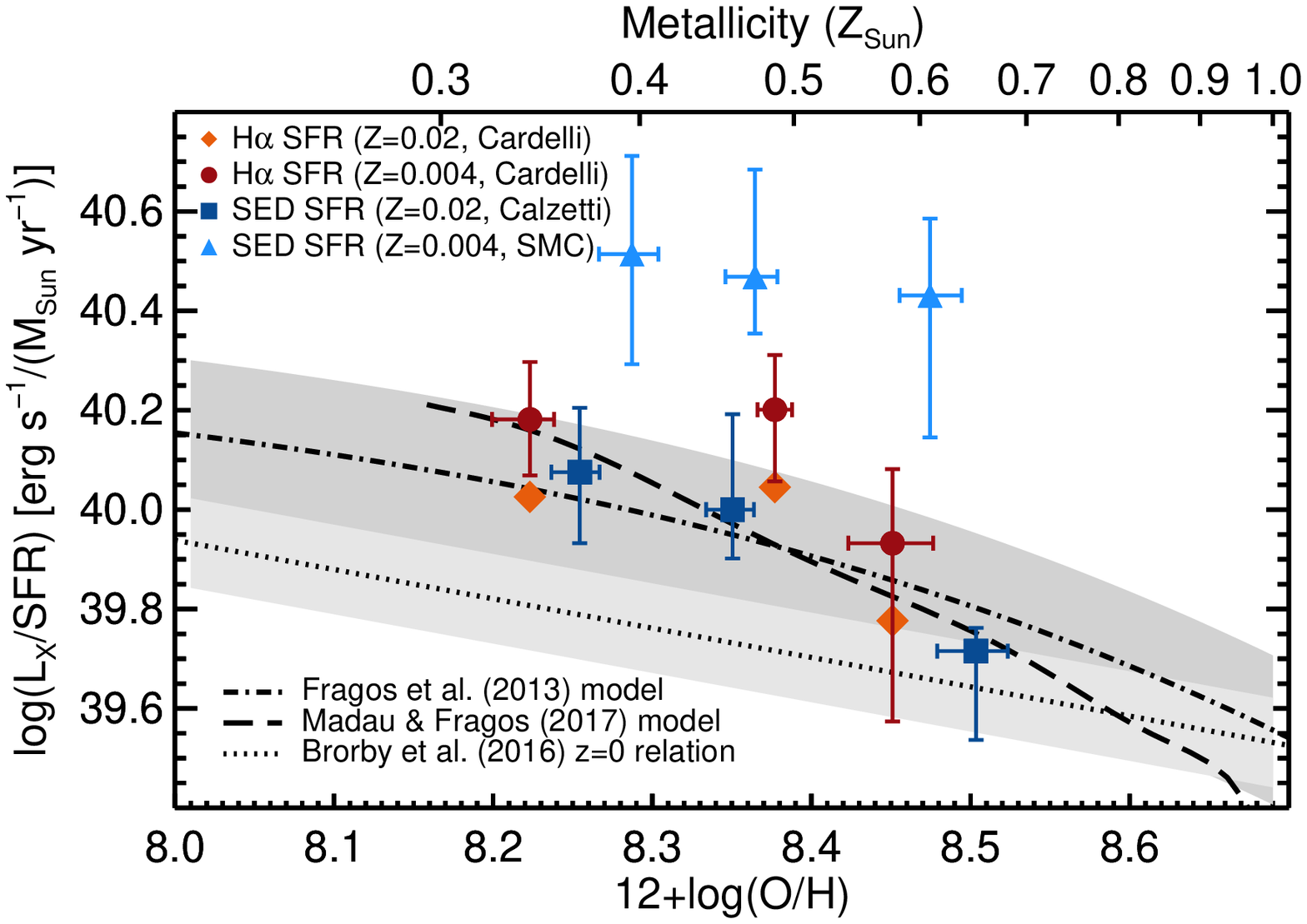}
\caption{$L_{\mathrm{X}}$/SFR versus stacked oxygen abundance for our galaxy sample restricted by sSFR.  The top axis shows the corresponding metallicity in solar units ($Z_{\odot}=0.0142$) assuming a solar abundance pattern.  Orange diamonds and red circles represent results based on H$\alpha$ SFRs assuming $Z=0.02$ and $Z=0.004$, respectively; the 1$\sigma$ error bars for these points are very similar, so only one set is shown for clarity.  Dark blue squares and light blue triangles represent results based on SED SFRs, assuming $Z=0.02$ and the \citet{calzetti00} attenuation curve for the former, and $Z=0.004$ and the SMC attenuation curve for the latter.  All SFR indicators favor an anti-correlation between $L_{\mathrm{X}}$/SFR and $Z$, but with different significance as discussed in \S\ref{sec:sfrindicator}.  The lines shown in this figure are as described in the caption for Figure \ref{fig:lxssfr}.}
\label{fig:lxsfrdiff}
\end{figure*}

A key source of systematic uncertainty which may impact our results is the choice of SFR indicator. Different SFR indicators probe different star formation timescales and it is debated how different indicators evolve with redshift.  As described in \S\ref{sec:hasfr}, for our default SFR measurements, we use H$\alpha$ SFRs with a $Z$-dependent $L_{\mathrm{H}\alpha}$-SFR conversion factor.  In this section, we explore how adopting different SFR indicators would impact our results.  \par
Figure \ref{fig:lxsfrdiff} displays $L_{\mathrm{X}}$/SFR as a function of $Z$ based on different SFR indicators.  Although we consider six different SFR indicators (see \S\ref{sec:sed}-\ref{sec:hasfr} for details), for simplicity we only show four of them in Figure \ref{fig:lxsfrdiff}, which provide a representative view of the impact of different SFR indicators.  Since SED-derived SFRs that adopt a delayed-$\tau$ SFH are consistent within 0.1 dex with those that adopt a constant SFH, we only discuss results based on the assumption of constant SFH.
 \par
The sSFR range of the galaxies in the stacks shown in Figure \ref{fig:lxsfrdiff} was restricted so that the $\langle sSFR \rangle$ of the different stacks varies by $<0.2$ dex.  As discussed in \S\ref{sec:zdep}, we found it is important to try to match the sSFR distribution between the different $Z$ stacks as much as possible to control for the sSFR-dependent LMXB contribution.  Since the typical SFRs of the galaxies in our sample vary depending on the SFR indicator, the common sSFR range we adopt also depends on the SFR indicator.  The log(sSFR/yr$^{-1}$) ranges are -9.2 to -8.4 for the $Z=0.02$ H$\alpha$ SFRs, -9.35 to -8.55 for the $Z=0.004$ H$\alpha$ SFRs, -9.05 to -8.5 for the $Z=0.02$ SED SFRs with the \citet{calzetti00} extinction curve, and -9.5 to -8.9 for the $Z=0.004$ SED SFRs with the SMC extinction curve. \par
For each SFR indicator, we calculate the probability that the stacked $L_{\mathrm{X}}$/SFR is anti-correlated with $Z$.  Assuming a power-law relationship between $L_{\mathrm{X}}$/SFR and $Z$, for all SFR indicators, we find that a negative power-law index is favored over an index $\geq 0$.  A negative correlation is favored with 83\% and 84\% probability when using the H$\alpha$-derived SFRs with $Z=0.02$ and $Z=0.004$, respectively, and with 99.4\% and 74\% confidence when using SED-derived SFRs with $Z=0.02$ and $Z=0.004$, respectively.  Thus, while regardless of SFR indicator, an anti-correlation between the $L_{\mathrm{X}}$/SFR of XRBs and $Z$ at z$\sim2$ is suggested by the data, the choice of SFR indicator does impact the significance of this result. \par
For each SFR indicator, we also calculate the high sSFR stacks as described in \S\ref{sec:hmxb}.  We find that, except for the stacks based on the SED-derived SFRs with $Z=0.004$, all the high sSFR stacks are in good agreement with the local $L_{\mathrm{X}}$-SFR-$Z$ relation from \citetalias{brorby16}, and they are inconsistent at $>99\%$ confidence with the \citetalias{madau17} model and the upper  $L_{\mathrm{X}}$/SFR bound of the \citetalias{fragos13b} models.  In particular, these results are not affected by the $L_{H\alpha}$-SFR conversion factor assumed.  The H$\alpha$ SFRs likely provide the most reliable comparison for the local $L_{\mathrm{X}}$-SFR-$Z$ relation, because the \citetalias{brorby16}, \citetalias{mineo12}, and \citetalias{lehmer10} local measurements are based on UV+IR SFRs and the H$\alpha$ SFRs are in good agreement with UV+IR SFRs for the MOSDEF sample \citep{shivaei16}.  Thus, the conclusion that the redshift evolution of $L_{\mathrm{X}}$/SFR for high sSFR galaxies is driven by the $Z$ dependence of HMXBs is fairly robust to the choice of SFR indicator.

\subsection{Other systematic effects}
\label{sec:systematics}

We investigate other sources of systematic uncertainty and their possible impact on our results.  

\subsubsection{Contamination from unidentified AGN}
While we have tried to screen out AGN as much as possible with multi-wavelength selection criteria, contamination from low-luminosity AGN remains a source of uncertainty.  \citet{fornasini18} find that even when luminous ($L_{\mathrm{X}}\gtrsim10^{42}-10^{43}$ erg s$^{-1}$) AGN are excluded, there is evidence for obscured, low-luminosity AGN with $\langle L_{\mathrm{X}} \rangle\approx10^{41}-10^{42}$ erg s$^{-1}$ in X-ray stacks of star-forming galaxies at $z\sim1-2$.  To gain some insight into how unidentified AGN may be influencing our results, we test how relaxing our AGN exclusion criteria impacts our stacks.  We experimented with including identified optical, IR, or X-ray AGN as well as all identified AGN to our stacks.  At most, this expands our galaxy sample by 34 galaxies and increases the $\langle L_{\mathrm{X}} \rangle$ of the middle-$Z$ and high-$Z$ stacks by 0.1 dex (the change in $\langle L_{\mathrm{X}} \rangle$ of the low-$Z$ stack is $<0.5$ dex).  While the resulting $L_{\mathrm{X}}$/SFR values of the stacks remain consistent within 1$\sigma$ statistical uncertainties, the inclusion of known AGN tends to flatten the observed $Z$ dependence.  The fact that $\langle L_{\mathrm{X}} \rangle$ does not significantly increase when galaxies above the \citet{kauffmann03} line are included in the X-ray stacks is consistent with observations that many normal star-forming galaxies at $z\sim2$ can lie above this line due to enhanced nebular N/O or stellar $\alpha$ enhancement, as compared with the ionized gas and massive stars in galaxies at $z=0$ (\citealt{masters14}; \citealt{shapley15}; \citealt{sanders16a}; \citealt{steidel16}).  The comparison of $L_{\mathrm{X}}$/SFR of stacks with and without identified AGN suggests that contamination from unidentified, low-luminosity AGN is unlikely to significantly impact the measured $L_{\mathrm{X}}$/SFR; if unidentified AGN have any impact at all, this comparison implies that the true HMXB-driven relation between $L_{\mathrm{X}}$/SFR and $Z$ may be even steeper than that observed.  Thus, the possibility of AGN contamination does not meaningfully impact our conclusions.   

\subsubsection{X-ray spectrum}
\begin{figure}
\centering
\includegraphics[width=0.5\textwidth]{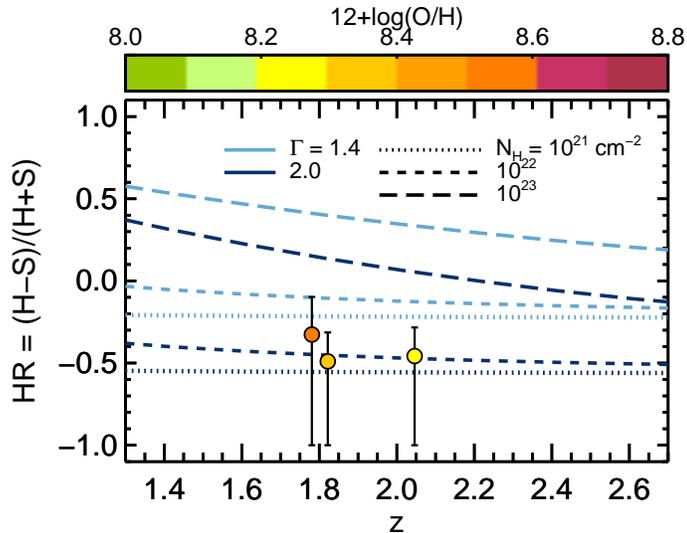}
\vspace{0.25in}
\caption{Hardness ratios of the three stacks based on the full sample versus weighted average redshift of the stack. Blue lines show the expected hardness ratios for sources with absorbed power-law spectra; line color represents different $\Gamma$ while line style represent different column densities ($N_{\mathrm{H}}$).  Our stacks are consistent with relatively unobscured ($N_{\mathrm{H}}<10^{22}$ cm$^{-2}$) spectra with $\Gamma=1.4-2.5$.  }
\label{fig:hratio}
\end{figure}

Another source of systematic uncertainty is the X-ray spectrum.  While our stacks do not have sufficient net counts for spectral fitting, hardness ratios can provide rough constraints on the spectrum.  For each stack, we calculated the hardness ratio based on the net counts in the $0.5-2$ keV (soft, S) and $2-7$ keV (hard, H) bands using the Bayesian estimation code BEHR \citep{park06}, which is designed for low count statistics. The hardness ratio is defined as $(H-S)/(H+S)$.  Figure \ref{fig:hratio} shows the hardness ratio of stacks \#4-6 (the hardness ratios of the other stacks are very similar).  As can be seen, these hardness ratios are consistent with relatively unobscured ($N_{\mathrm{H}}\lesssim10^{22}$ cm$^{-2}$) spectra with a photon index of $\Gamma=1.4-2.5$, and our adopted spectral model (unobscured, $\Gamma=2$ power-law) falls within this range.  Varying the spectral parameters within the allowed ranges results in $+/-0.15$ dex variations in the stacked $L_{\mathrm{X}}$, which is comparable to the statistical uncertainties; thus, the general agreement of our stacks with the \citetalias{fragos13b} model and the local relation is not substantially affected by spectral variations.  Changing the spectral parameters affects all stacks by the same logarithmic amount, so the relative differences between the stacks remain unchanged.  
      
\subsubsection{Redshift evolution of $Z$ indicator and chemical abundances}
Even though we use the same $Z$ indicator as \citet{brorby16} to facilitate the comparison of the MOSDEF $z\sim2$ stacks with the local $L_{\mathrm{X}}$-SFR-$Z$ relation, it remains debated how O3N2 (and other $Z$ indicators) may evolve with redshift.  Studies of emission line ratios at $z\sim2$ find that the N2 indicator overestimates the gas-phase O/H due to either elevated N/O or a harder stellar ionizing spectrum at fixed O/H (\citealt{masters14}; \citealt{shapley15}; \citealt{steidel16}).  While these factors may also affect the O3N2 indicator, \citeauthor{steidel14} (\citeyear{steidel14}; \citeyear{steidel16}) present evidence that the O3N2 indicator is not significantly biased.  While oxygen based indicators such as O$_{32}$ or R$_{23}$ are coming to be considered more reliable than indicators which include nitrogen (\citealt{shapley15}; \citealt{sanders16a}; \citealt{sanders16b}), adopting one of these indicators would decrease our sample size even further.  Furthermore, O$_{32}$ and R$_{23}$ are more impacted by reddening \citep{sanders18}.  Comparing the oxygen abundances derived using O3N2 and O$_{32}$ for the subset of MOSDEF galaxies for which both indicators are available, we find that O/H values based on O3N2 are on average 0.13 dex lower than those derived from O$_{32}$.  If the $\langle$ O/H $\rangle$ values of our stacks are systematically offset by this amount, they would remain in agreement with the local $L_{\mathrm{X}}$-SFR-$Z$ relation, but some of the tension with the \citetalias{madau17} and upper \citetalias{fragos13b} models would be eased.
\par
Finally, especially when comparing our results to theoretical models, it is important to keep in mind that there may be differences between nebular O/H, which we use as a $Z$ proxy, and stellar metallicity as defined by the \citetalias{fragos13b} and \citetalias{madau17} models.  In particular, the chemical abundances in $z\sim2$ galaxies may be different from the solar abundance pattern assumed by the models.  The $Z$ dependence of radiatively-driven stellar winds, which is the underlying cause of the $Z$ dependence of $L_{\mathrm{X}}$/SFR in HMXB population synthesis models, primarily depends on the abundance of Fe in the case of solar abundance ratios \citep{vink01}.  However, \citet{steidel16} found that $z\sim2$ star-forming galaxies can be highly supersolar in O/Fe, as expected for a gas that is primarily enriched by core collapse supernovae.  
Using a small sample of $z\sim2$ galaxies with [O{\small III}]$\lambda4363$ detections, including four MOSDEF galaxies, \citet{sanders19} similarly find that O/Fe is enhanced in the galaxies, but note that neither their sample nor the \citet{steidel16} sample may be representative of $z\sim2$ galaxies. \par
Nonetheless, let us consider the implications if the MOSDEF $z\sim2$ galaxies are typically supersolar in O/Fe.  In this case, the line-driven winds of their stellar populations are likely dominated by C, N, and O rather than Fe.  While the results of \citet{vink01} suggest that the $Z$ dependence of Fe-driven and CNO-driven winds may be similar, this issue has not been investigated for $Z>0.1 Z_{\odot}$.  The \citetalias{fragos13b} and \citetalias{madau17} models we use as a point of comparison, like most current models, assume solar abundance ratios for $Z\gtrsim0.1 Z_{\odot}$, and thus their appropriateness for high redshift stellar populations should be investigated. \par
In summary, none of these systematic effects substantially alter the conclusion that the observed redshift evolution of $L_{\mathrm{X}}$/SFR is consistent with being driven by the $Z$ dependence of HMXBs.

\section{Conclusions}
\label{sec:conclusions}

We have studied the X-ray emission of a sample of 79 star-forming galaxies at $1.37<z<2.61$ in the CANDELS fields with rest-frame optical spectra from the MOSDEF survey in order to investigate the 
metallicity dependence of HMXBs at $z\sim2$.  While studies of local galaxies have discovered that HMXB populations in low-$Z$ galaxies are more luminous (e.g. \citealt{brorby16}), and the observed increase of $L_{\mathrm{X}}$/SFR with redshift has been attributed to this $Z$ dependence (\citealt{basu13a}; \citealt{lehmer16}), the connection between the redshift evolution and $Z$ dependence of HMXBs has not been directly tested previously.  
In order to assess whether the $Z$ dependence of HMXBs can account for the observed increase in $L_{\mathrm{X}}$/SFR as a function of redshift, we (a) tested whether the $L_{\mathrm{X}}$/SFR of HMXBs depends on $Z$ at $z\sim2$ and (b) compared this trend to the local $L_{\mathrm{X}}$-SFR-$Z$ relation.
\par
After removing AGN based on multi-wavelength diagnostics, we stacked the X-ray data of star-forming galaxies from the \textit{Chandra} AEGIS-X Deep survey, the \textit{Chandra} Deep Field North, and the \textit{Chandra} Deep Field South.  Investigating how the $L_{\mathrm{X}}$/SFR of our galaxies varies when they are grouped according to redshift, $Z$, and sSFR, we find the following results:
\begin{enumerate}
\item The average $L_{\mathrm{X}}$/SFR of galaxies at $z\sim1.5$ and $z\sim2.3$ is elevated compared to values for local star-forming galaxies (\citealt{lehmer10}; \citealt{mineo12}).  This $\approx0.4-0.8$ dex enhancement is comparable to that observed for $z\sim2$ galaxies by previous studies (\citealt{lehmer16}; \citealt{aird17}).  
\item Splitting our sample into three metallicity bins, we find that $L_{\mathrm{X}}$/SFR and $Z$ are anti-correlated with 97\% confidence at similar sSFR.  This result is based on H$\alpha$-derived SFRs with $Z$-dependent conversion factors, which we consider to be the most reliable SFR indicator available for this galaxy sample.  It provides the first evidence for the metallicity dependence of XRB populations at $z>0$.  This trend is more likely to be driven by HMXBs than LMXBs, unless $L_{\mathrm{LMXB}}/M_*$ decreases by a factor of $\approx7$ as 12+log(O/H) increases from 8.25 to 8.45.  Such large variation would be challenging to explain using current population synthesis models.
\item Stacking only galaxies with high sSFR (sSFR$>1.6\times10^{-9}$ yr$^{-1}$) in order to minimize the contribution from LMXBs, we find that the $L_{\mathrm{X}}$/SFR values of our sample are consistent with the local $L_{\mathrm{X}}$-SFR-$Z$ relation \citep{brorby16}.  Thus, HMXB populations at $z\sim2$ lie on the same $L_{\mathrm{X}}$-SFR-$Z$ relation as galaxies at $z=0$.  The high sSFR stacks disagree at $>3\sigma$ confidence with the upper $L_{\mathrm{X}}$/SFR bound of the \citetalias{fragos13b} HMXB models and the best-fit HMXB population synthesis model from \citetalias{madau17}.  
\end{enumerate}
The three preceding results combined provide direct evidence that the enhanced $L_{\mathrm{X}}$/SFR of $z\sim2$ star-forming galaxies compared to high sSFR galaxies of similar $M_*$ in the local Universe is due to the lower metallicity of the HMXB populations in high redshift galaxies.  This study thus supports the hypothesis of previous works (\citealt{basu13a}; \citealt{lehmer16}) that the observed redshift evolution of $L_{\mathrm{X}}$/SFR is the result of the $Z$ dependence of HMXBs combined with the fact that higher-redshift galaxy samples have lower metallicities on average.  \par
By comparing stacks with different sSFR but similar $Z$, we are also able to estimate that the $L_{\mathrm{X}}/M_*$ due to LMXBs is $2-8\times10^{30}$ erg s$^{-1}$ $M_{\odot}^{-1}$ at $z\sim2$.  This estimate is an an order of magnitude higher than local values (\citealt{gilfanov04}; \citealt{lehmer10}, but consistent with predictions from the \citetalias{fragos13b} and \citetalias{madau17} LMXB population synthesis models. \par
Possible AGN contamination, the assumed X-ray spectrum, and systematics associated with the metallicity measurements do not significantly impact our conclusions.  The choice of SFR indicator can substantially affect the absolute $L_{\mathrm{X}}$/SFR values, but the result that $L_{\mathrm{X}}$/SFR varies with $Z$ and that this trend is consistent with the $L_{\mathrm{X}}$-SFR-$Z$ local relation are fairly robust to the choice of SFR indicator.  
As our understanding of SFR indicators at high redshift improves, it will be important to revisit these issues.  \par
Furthermore, since there is evidence that the stellar populations at $z\sim2$ may have supersolar O/Fe, it is also important to investigate the effect of $\alpha$-element enhanced abundances on HMXB population synthesis models.  Current models assume solar abundance ratios, and have not studied the impact of CNO rather than Fe driven stellar winds on HMXB populations with $Z>0.1Z_{\odot}$.
\par
While our study only probes the metallicity range of 12+log(O/H)$=8.0-8.8$, our results indicate that the \citetalias{brorby16} relation and the \citetalias{fragos13b} models with lower $L_{\mathrm{X}}$/SFR normalizations provide reasonable estimates of the X-ray emission of HMXBs out to high-redshift.  Thus, we encourage the adoption of these scaling relations by studies searching for faint X-ray AGN or investigating the effect of X-ray heating on the epoch of reionization.
\par
While this study provides the first direct connection between the redshift evolution and $Z$ dependence of HMXBs, future work is required to improve the statistical significance of this result and to constrain theoretical models of the $Z$ dependence of HMXBs.  Larger samples of galaxies with $Z$ measurements are crucial to reduce the statistical uncertainties of the stacked  $L_{\mathrm{X}}$.  Future thirty-meter class telescopes will be critical for increasing such measurements at high redshifts.  Expanding this study to other redshift ranges will also further test 
whether the observed redshift evolution is driven by the HMXB $Z$ dependence as found by this study.  Furthermore, improving measurements of the local $L_{\mathrm{X}}$-SFR-$Z$ relation by increasing the local galaxy sample size and determining its dependence on additional variables such as sSFR is important as it provides a benchmark for high-redshift studies.  \par
With current X-ray instruments, the scatter in the $L_{\mathrm{X}}$-SFR-$Z$ relation can only be studied using nearby galaxy samples, while higher-redshift studies depend on stacking or other statistical techniques.  In the future, the \textit{Athena X-ray Observatory} will enable the detection of  large samples of individual XRB-dominated galaxies out to $z\sim1$, and the \textit{Lynx X-ray Observatory} would push these detection limits out to $z\sim6$.  Combined with accurate $Z$ and SFR measurements from \textit{JWST}, the large samples of individually detected XRB-dominated galaxies provided by these future X-ray missions will enable much more detailed investigations of the multivariate dependence of $L_{\mathrm{XRB}}$ and its scatter on galaxy properties and redshift.  These future analyses will help provide stronger constraints on models of stellar evolution, the progenitor channels of gravitational wave sources, and the X-ray heating of the intergalactic medium in the early Universe.  

\acknowledgments

We thank G. Fabbiano and M. Elvis for fruitful conversations about this study.  FMF and MK acknowledge support from \textit{Chandra} grant 17620679.  We acknowledge support from NSF AAG grants AST1312780, 1312547, 1312764, and 1313171, archival grant AR13907 provided by NASA through the Space Telescope Science Institute, and grant NNX16AF54G from the NASA ADAP program. We additionally acknowledge the 3D-HST collaboration for providing spectroscopic and photometric catalogs used in the MOSDEF survey.  The scientific results reported in this paper
are based on observations made by the \textit{Chandra X-ray Observatory}. This study also made use of data obtained at the W.M. Keck Observatory, which is operated as a scientific partnership among the California Institute of Technology, the University of California, and the National Aeronautics and Space Administration. The Observatory was made possible by the generous financial support of the W.M. Keck Foundation.  We wish to extend special thanks to those of Hawaiian ancestry, on whose sacred mountain we are privileged to be guests.  Without their generous hospitality, this work would not have been possible.  This study is also based on observations made with the NASA/ESA Hubble Space Telescope, which is operated by the Association of Universities for Research in Astronomy, Inc., under NASA contract NAS 5-26555. 

{\large \textit{Software}:} CIAO \citep{fruscione06}, BEHR \citep{park06}

\facilities{Chandra X-ray Observatory, Keck Observatory}

\vspace{5mm}

\bibliographystyle{aasjournal}
\bibliography{refs}

\end{document}